\documentclass[superscriptaddress,aps,amsfonts,notitlepage,onecolumn,nofootinbib]{revtex4-2}

\usepackage{xcolor,graphicx}
\usepackage{breqn,amsmath,amssymb}%for automatic breaking of equation in\begin{dmath} and tensor indices
\usepackage{mathtools}
\usepackage{enumerate}

\usepackage{booktabs}%to merge cells 
\usepackage{makecell}% to split text inside a cell

\usepackage{comment}

\usepackage{subcaption}

\usepackage[T1]{fontenc}
\usepackage{natbib,hyperref}

\usepackage{mathtools}

\usepackage{centernot}

\usepackage[utf8]{inputenc}

\definecolor{linkcolor}{HTML}{799B03}
\definecolor{urlcolor}{HTML}{799B03}
\usepackage[english]{babel}
\usepackage{amsmath, amsthm, amsfonts, amssymb,amscd}
\usepackage[all]{xy}
\usepackage{graphicx}
\usepackage{import}
\usepackage{xifthen}
\usepackage{float}
\usepackage{transparent}
\usepackage{multirow}
\usepackage{cellspace}
\usepackage{colortbl}
\usepackage{bbm}
\usepackage{mathrsfs}
\usepackage{csquotes}
\usepackage{natbib}

\graphicspath{{pic/}}
\def\[{\begin{equation}}
\def\]{\end{equation}}
\def\e{\mathrm{e}}
\def\i{\mathrm{i}}

\DeclareUnicodeCharacter{0301}{`}

\makeatletter%These 3 lines to avoid conflicts between package breqn that automatically breaks long equations and putting comma inside affiliations
\let\cat@comma@active\@empty
\makeatother

\usepackage{IEEEtrantools}

\begin{document}

%\bstctlcite{IEEEexample:BSTcontrol}
\title{Higher derivative SVT theories from Kaluza-Klein reductions of Horndeski theory}

\author{S. Mironov}
\email{sa.mironov\_1@physics.msu.ru}
\affiliation{Institute for Nuclear Research of the Russian Academy of Sciences, 
60th October Anniversary Prospect, 7a, 117312 Moscow, Russia}
\affiliation{Institute for Theoretical and Mathematical Physics,
MSU, 119991 Moscow, Russia}
\affiliation{NRC, "Kurchatov Institute", 123182, Moscow, Russia}

\author{A. Shtennikova}
\email{shtennikova@inr.ru}
\affiliation{Institute for Nuclear Research of the Russian Academy of Sciences, 
60th October Anniversary Prospect, 7a, 117312 Moscow, Russia}
\affiliation{Institute for Theoretical and Mathematical Physics,
MSU, 119991 Moscow, Russia}
%\affiliation{Department of Particle Physics and Cosmology, Physics Faculty, M.V. Lomonosov Moscow State University,
%Vorobjevy Gory, 119991 Moscow, Russia}

\author{M. Valencia-Villegas}
\email{mvalenciavillegas@itmp.msu.ru}
\affiliation{Institute for Theoretical and Mathematical Physics,
MSU, 119991 Moscow, Russia}
\affiliation{Institute for Nuclear Research of the Russian Academy of Sciences, 
60th October Anniversary Prospect, 7a, 117312 Moscow, Russia}

\begin{abstract}
It was recently pointed out that some precise Photon-Galileon couplings in four dimensions (4D) -inspired by a higher dimensional reduction- are enough to obtain a Horndeski theory that is less constrained by the stringent experimental bounds on the speed of Gravitational Waves. They imply the constancy of the ratio of speed of gravity to light throughout cosmic evolution. This holds even if we include the general scalar potentials $G_4 (\pi,X)$ and $G_5 (\pi)$. In this paper we go into the details of this 4D Luminal extension of Horndeski theory including its scalar sector. We also present the complete action including the general $G_5(\pi,X),\, G_6(\pi,X)$ scalar potentials. Thus we show all the $U(1)$ gauge invariant vector Galileons  in 4D that result from a Kaluza-Klein dimensional reduction from 5D Horndeski. They provide a {\it consistent} coupling of a higher derivative vector to scalar modifications of gravity --- namely, without inducing Ostrogradsky ghosts and keeping gauge invariance--- in the aim to explore more universal couplings of dark energy to other matter, such as vectors and in particular the Photon.
\end{abstract}

\maketitle
\tableofcontents

\section{Introduction}

The strict experimental bounds on the difference between the speed of gravitational and electromagnetic waves, deduced from the event GW170817 \cite{LIGOScientific:2017vwq}, heavily constrained the application of Horndeski theory/ Generalized Galileons --- the most general modification of GR with a higher derivative scalar in the action, but with second order equations of motion\footnote{In this work we are only concerned with the classical solutions and thus, it is relevant that the equations of motion are of second order. In particular we do not discuss quantum corrections and the higher derivative operators that could become relevant in some specific physical situation.  Nevertheless, it would be interesting to assess the quantum robustness of the aspects discussed in this work.} \cite{horndeski1974second,nicolis2009galileon,Deffayet:2011gz}--- for late time cosmology \cite{LIGOScientific:2017vwq,Abdalla:2022yfr, Bettoni:2016mij,Ezquiaga:2017ekz,Sakstein:2017xjx,Baker:2017hug,Creminelli:2017sry}. The standard approach to deal with these experimental constraints has been simply pragmatical: take advantage of the generality of Horndeski theory by choosing the four scalar potentials in the action to satisfy the sub/ luminality of the graviton (and other physically compelling conditions in cosmological solutions). Nevertheless, it was the main point in a recent work \cite{Mironov:2024idn} that those constraints on Horndeski theory assumed that the well known Photon of Maxwell electrodynamics {\it remains without modifications} in the IR scale of gravity, and that lifting this assumption offers a new range of possibilities. Indeed, it was shown in \cite{Mironov:2024idn} that there can be Photon-Galileon couplings that can account for the {\it exact} constancy of the ratio of speed of gravity to light throughout cosmic evolution, even with the general potentials $G_4 (\pi,X)$ and $G_5 (\pi)$ \footnote{A {\it part} of the result in \cite{Mironov:2024idn} --- whose scalar sector we show below --- relies on the assumption that the scalar potential $G_5$ depends only on $\pi$ and not on $X$. However, one must be aware that in some physical situations where quantum corrections are relevant, even if one assumes a $X$ independent $G_5(\pi)$, quantum corrections will generate  the $X$ dependence in the absence of a symmetry that avoids it. See for instance \cite{Pirtskhalava:2015nla, Goon:2016ihr}.}. Furthermore, it was also recently shown  for the latter class that the same luminal property holds on more general spherically symmetric and dynamical backgrounds \cite{MironovNew,Mironov:2024yqa}. Therefore, the latter could be relevant to promising proposals in Galileons, for instance, the {\it Fab Four} to address the cosmological constant problem \cite{Charmousis:2011bf} and which were strictly constrained assuming the minimally coupled Photon \cite{Ezquiaga:2017ekz}. Now, the question of how to recover the usual Maxwell Photon in laboratory tests from that proposed IR modification was also briefly discussed in \cite{Mironov:2024idn}, and a recent analysis in Beyond Horndeski theory has also touched on the topic of experimental constraints on the new type of Photon-Galileon couplings proposed in \cite{Mironov:2024idn} and the analogous in the context of  Beyond Horndeski \cite{Babichev:2024kfo}.\\

Thus, the exploration of for instance Photon-(Dark Energy) couplings and more broadly Matter-(Dark Energy) couplings--- if we use Galileon models to try to account for Dark Energy ---  can be an important new way to pick up experimentally favored directions in modified gravity for late time cosmology using laboratory tests. Indeed, the universal coupling of gravity to all forms of energy may be a motivation for the exploration of more ubiquitous couplings of IR modifications of gravity to other energy sources, specially given the possibility to account for the speed tests of GW and light with the latter \cite{Mironov:2024idn}. In this work we expand on the results shown in the recent letter \cite{Mironov:2024idn} without restricting to the luminal cases. We use the same analysis to show the dimensional reduction of Horndeski theory in five dimensions ($5D$) to obtain a broad class of $U(1)$ gauge invariant vector/ multi-scalar Galileons in 4D: namely, a theory with higher derivatives of a $U(1)$ gauge invariant vector and two scalars in the action, but with second order equations of motion for all of the fields. Such types of theories had been previously obtained from different considerations \cite{Horndeski:1976gi} and also by Kaluza-Klein (KK) reductions from, for instance, Gauss-Bonnet in 5D \cite{Buchdahl:1979wi,Mueller-Hoissen:1987nvb}. Also, more recently, Horndeski Scalar-Vector couplings were obtained from KK reductions of Lovelock theory \cite{Charmousis:2008kc}, and bi-scalar theories in $2D$ were obtained from $4D$ Horndeski \cite{Nejati:2024tuo}. However, given that we start from the most general scalar modification of Lovelock theory in 5D with higher derivatives in the action, but with second order equations of motion --- Horndeski/ Galileons in 5D---, we obtain a much {\it larger class} of Vector-Scalar couplings in 4D, while obviously preserving the second order equations of motion. Indeed, since the 5D metric solves second order equations, its components (4D metric, vector and Dilaton) solve also second order equations as well as the Galileon, by construction. The task of building such gauge invariant Vector Galileons and Einstein-Maxwell Lagrangians in a systematic way is daunting from other perspectives \cite{Tasinato:2013oja,Petrov:2018xtx,Deffayet:2013tca,Colleaux:2023cqu,Colleaux:2024ndy}, and hence the interest of using the dimensional reduction also as a tool to expand the theory space of Vector-Scalar couplings while keeping gauge invariance and avoiding the Ostrogradsky ghost. We, nevertheless, do not aim for completeness, nor for a complete classification, but rather put emphasis on these theories with the interest that first, a subclass of it already leads to a luminal graviton in 4D (for very general scalar potentials and nonminimal derivative couplings to gravity) \cite{Mironov:2024idn}; and second, because being connected to higher dimensional theories\footnote{An additional interest on this regard comes from the very origin of Galileons, which were inspired by the  effective theory of the DGP model \cite{nicolis2009galileon} and further relations that have been established with higher dimensional theories \cite{VanAcoleyen:2011mj,deRham:2010eu,Trodden:2011xh}, and to string theory \cite{Easson:2020bgk}.} they could be simpler to analyze in some cases, for instance, in the look for Black-Hole solutions in 4D \cite{Charmousis:2008kc,Charmousis:2014mia} (See for instance the connection of covariant Galileons in general D dimensions \cite{Deffayet:2009mn} --- a special case of our starting point, 5D generalized Galileons --- to Lovelock theory in yet higher $D+N$ dimensions by KK diagonal compactifications  \cite{VanAcoleyen:2011mj}). Furthermore, vectors have been long studied for cosmological applications \cite{Esposito-Farese:2009wbc,Golovnev:2008cf,BeltranJimenez:2008iye}, and so the consistent coupling to scalar modifications of gravity --- without inducing ghosts, and keeping gauge invariance---  could shed light on promising theories, however, their application is beyond the scope of this work. 

A derivation of these $U(1)$ gauge invariant vector/ multi-scalar Galileons in 4D, which, as is also common, we denote as SVT (Scalar-Vector-Tensor theories), is given in section \ref{sec vector galileons}.\\

On the other hand, given the motivation of these SVT theories for IR modifications of gravity by {\it a scalar field}, but also {\it more ubiquitously coupled to other matter} such as the Photon, we start the discussion on the stability of nonsingular cosmological solutions. Indeed, modified theories of gravity such as Horndeski theory can violate the Null-Energy-Condition (NEC) without obvious pathologies. Thus, this compelling feature has been widely used to build interesting cosmological solutions not possible in GR, such as nonsingular universes, among others. However, it has been also shown in the form of a No-Go theorem \cite{Libanov:2016kfc, Kobayashi:2016xpl} that they suffer, in general, of global stability issues \cite{Evslin:2011vh,Easson:2011zy,Sawicki:2012pz,Rubakov:2016zah,Kolevatov:2016ppi,Mironov:2019fop,Akama:2017jsa,Cai:2017dyi,Creminelli:2016zwa,Cai:2016thi,Cai:2017tku,Kolevatov:2017voe}. More specifically, even if far from the most physically relevant phase of cosmic evolution, instabilities will arise earlier or later in any nonsingular, nonghosty cosmological solution\footnote{This holds at first order and assuming conventional asymptotics in the general case. See also \cite{Cai:2016thi,Cai:2017tku,Creminelli:2016zwa,Cai:2017dyi,Kobayashi:2016xpl,Ageeva:2021yik,Mironov:2022quk}.}. This No-Go was even generalized to multi-Galileons \cite{Akama:2017jsa} and to a subclass of  Horndeski theory on a spacetime with torsion \cite{Mironov:2023wxn}, although it was also shown to be broken for the full Horndeski theory with both curvature and torsion \cite{Mironov:2024ffx}. It is important to mention that an alternative to avoid the No-Go is to consider Beyond-Horndeski theory \cite{Cai:2017dyi,Creminelli:2016zwa,Cai:2016thi,Kolevatov:2017voe}. However, in this work we stay within the Horndeski class, namely, with second order equations of motion and we use the No-Go theorem in multi-Galileons to show that the theories considered here also suffer these global instability issues. In particular, this result applies to the theories with a Luminal graviton that were proposed in \cite{Mironov:2024idn}.\\

In section \ref{sec linearization} we introduce notation for the first order perturbation on the spatially flat FLRW cosmological background. In section \ref{sec luminality} we compute the quadratic action for the tensor and vector sectors and review the cases in which the automatical luminality of GWs is achieved. In section \ref{sec screening} ---restricted to  these luminal cases--- we also discuss  the identification of the vector mode  with the Photon of Maxwell electrodynamics in the same line of reasoning that it is common practice to identify the tensor modes of Horndeski with the graviton of GR.

In section \ref{sec stability} we show the scalar modes and discuss the stability issues of nonsingular cosmological solutions. We end with conclusions in section \ref{sec conclusions}.

\section{Obtaining 4D Luminal Horndeski from a higher dimensional reduction}\label{sec vector galileons}

The standard formulation of Horndeski theory/ Generalized Galileons in 4D assumes that the Photon is minimally coupled to gravity. Namely, the scalar field (Galileon) couples to curvature and it is assumed that the electromagnetic 4-vector does not couple to the Galileon. Thus, the speed of the photon $c_A$ is a constant throughout cosmic evolution ($1$ in our units). In \cite{Mironov:2024idn} a new type of Photon-Galileon couplings was found --- inspired by a KK  reduction of a higher dimensional scalar  tensor theory --- in such a way that {\it what is kept constant in time} is not the speed of light but {\it the ratio of speeds} of gravity $c_g$ and light, namely
\begin{equation}
    \frac{c_A(t)}{c_g(t)}=1\,,\label{eqn ratio}
\end{equation}
while $c_A(t)$ and $c_g(t)$ can vary on the {\it cosmological} background.
Indeed, it seems clear -up to a caveat to be shown below- that if the graviton and photon in 4D are different components of the same "gravitational wave" in 5D, then the speed of gravity and light will coincide in 4D. This turns out to be the case for a large class -{\it but not all}- of Horndeski theories in 4D with Photon-Scalar couplings derived by the KK dimensional reduction. Thus, the KK reduction is a {\it useful tool} to find theories with the property (\ref{eqn ratio}). The latter were the main topic in \cite{Mironov:2024idn} and in this work we cover in more detail both of these types theories, those with and without a luminal graviton.

For completeness, let us review the idea proposed in \cite{Mironov:2024idn}. Let us consider Generalized Galileons formulated in 5D. It is then clear that we do not aim to compare nor discuss the fundamental status of the 5D theory against the 4D theory. Both of them are simply  modifications of gravity, but as we will see, with interesting properties in 4D.

Denoting all quantities defined on the 5D manifold $M_5$ with hats, such as $\hat{R}$ -curvature tensors computed with the 5D metric $\hat{g}$ (with mostly minus signature)-, and choosing indices that are capital Latin letters running over the five dimensions, $M = 0, 1, 2, 3, 4$, while lowercase Greek letters run over our four dimensions, $\alpha = 0, 1, 2, 3$, the 5D Galileons can be written as \cite{Deffayet:2011gz}

\begin{subequations}
    \label{lagrangian}
    \[S=\int\mathrm{d}^5x\sqrt{\hat{g}}\,\mathcal{L}_\pi,\]
    \[\mathcal{L}_\pi=\mathcal{L}_2 + \mathcal{L}_3 + \mathcal{L}_4 + \mathcal{L}_5 + \mathcal{L}_6\]
    \vspace{-0.8cm}
    \begin{align}
    &\mathcal{L}_2=F(\pi,X),\\
    &\mathcal{L}_3=K(\pi,X)\hat{\Box}\pi,\\
    &\mathcal{L}_4=-G_4(\pi,X)\hat{R}+2G_{4X}(\pi,X)\left[\left(\hat{\Box}\pi\right)^2-\pi_{;MN}\pi^{;MN}\right],\\
    &\mathcal{L}_5=G_5(\pi,X)\hat{G}^{MN}\pi_{;MN}+\frac{1}{3}G_{5X}(\pi,X)\left[\left(\hat{\Box}\pi\right)^3-3\hat{\Box}\pi\pi_{;MN}\pi^{;MN}+2\pi_{;MN}\pi^{;MP}\pi_{;P}^{\;\;N}\right],&\\
    &\mathcal{L}_{6} = \dfrac{3}{4}{G_{6}(\pi,X)} \left({\hat{R}}^{2}-4\hat{R}^{A B} \hat{R}_{A B}+\hat{R}^{A B C D} \hat{R}_{ABCD}\right)+3\, G_{6X}(\pi,X) \left(-\hat{R} \left({\left(\hat{\Box}\pi\right)}^{2}-\pi^{;AB}\pi_{;AB}\right)\right.&\nonumber\\
    &\left.+4 \hat{R}^{AB} \hat{\Box}\pi\, \pi_{;AB}-4 \hat{R}^{AB} \pi_{;A}\,^{C} \pi_{;CB}-2\hat{R}^{ABCD} \pi_{;AC} \pi_{;BD}\right) + G_{6XX}(\pi,X) \left({\left(\hat{\Box}\pi\right)}^{4}\right.&\nonumber\\
    &\left.-6\,\pi^{;AB}\pi_{;AB} {\left(\hat{\Box}\pi\right)}^{2}+8\,\hat{\Box}\pi\, \pi^{;AB} \pi_{;B}\,^{C} \pi_{;CA}+3\left({\pi^{;AB}\pi_{;AB}}\right)^{2}-6\pi^{;AB} \pi_{;B}\,^{C} \pi_{;C}\,^{D} \pi_{;DA}\right),&
    \end{align}
\end{subequations}
where $\pi$ is the scalar field (Galileon), $X=\hat{g}^{M N}\pi_{;M}\pi_{;N}$, $\pi_{;M}=\partial_M\pi$, $F,\, K,\, G_4,\, G_5,\, G_6$ are general scalar potentials, $\pi_{;MN}=\hat\nabla_N \hat\nabla_M\pi$, $\hat\Box\pi = \hat{g}^{MN}\hat\nabla_N\hat\nabla_M\pi$, and we have defined as $G_{4X}=\partial G_4/\partial X$ and so on.\\

Now, let us consider the 5D Galileons (\ref{lagrangian}) compactified on a circle. We write the metric as usual in terms of $h_{\mu\nu}$, a 4-vector $A_\mu$ and a Dilaton $\phi$, assuming the "cylinder condition" \footnote{We assume that $M_5$ can be decomposed as a direct product of a pseudo-Riemannian manifold, $M_4$, and the circle, $S^1$, such that $M_5 = M_4 \times S^1$. This decomposition allows us to expand fields on $M_5$ as Fourier series in the additional coordinate, $x^4$. Specifically, we have:
\[ \rho = \sum_n \rho_n \e^{\i n x_4} \]
where $\rho$ represents any field on $M_5$, and the coefficients $\rho_n$ are functions of the coordinates $x^\mu$ on $M^4$. If there are any terms in the series other than $\rho_0$, the fields defined on $M_5$ would give rise to an infinite number of massive fields after reduction on $M^4$. To avoid this, we impose the condition that all quantities are independent of the fifth dimension, namely $\rho = \rho(x^\mu) $.} \cite{Kaluza:1921tu},
\begin{eqnarray}\label{metric}
  \hat{g}_{MN} = \begin{pmatrix}
   h_{\mu \nu} - \phi^2 A_{\mu} A_{\nu}  & \phi^2 A_{\mu}\\
    \phi^2 A_{\nu} & -\phi^2
\end{pmatrix},
\end{eqnarray}
and with inverse,
\begin{equation}
  \hat{g}^{MN} = \begin{pmatrix}
 h^{\mu \nu}   &  h^{\mu \nu} A_{\nu}\\
     h^{\mu \nu} A_{\mu} &  -\frac{1}{\phi^2} + h^{\mu\nu}A_{\mu} A_{\nu}
\end{pmatrix}.
\end{equation}
As usual we identify $h_{\mu\nu}$ as the metric on our 4D spacetime. Now, at least in the Luminal theories shown in \cite{Mironov:2024idn} --- considered as IR modifications of gravity --- we will identify $A_\mu$ with the Photon because, first, it is a $U(1)$ gauge invariant theory in 4D and second, because the vector modes propagate at exactly the same speed as the tensor modes of Horndeski gravity, which is suggested by recent observations \cite{Bettoni:2016mij,LIGOScientific:2017vwq,Ezquiaga:2017ekz,Abdalla:2022yfr}. This is further discussed in section \ref{sec screening}, although this identification is irrelevant for the most part of this work.

With the metric (\ref{metric}) and the cylinder condition, normalizing our fields to reabsorb the multiplicative constant in the action\footnote{Namely, in the dimensional reduction we integrate the action~\eqref{lagrangian} on the coordinate $x^4$ because all fields of the theory do not depend on this coordinate.} $\int dx^4$, we rewrite the 5D Galileon (\ref{lagrangian}) as a 4D theory (See the Appendix \ref{sec decomp} for details) 
\begin{equation}
    S=\int\mathrm{d}^4x\sqrt{-h}\,\phi\,\left(\mathcal{L}_\pi+\mathcal{L}_{\phi} + \mathcal{L}_{A}\right)\,,\label{lagrangian_after_reduction}
\end{equation}
where $\mathcal{L}_\pi$ takes the form of (\ref{lagrangian}), but removing the hat of all the quantities, namely, computing all curvature tensors and covariant derivatives with the 4D metric $h$ (the decomposition of five-dimensional scalar structures in terms of four-dimensional ones can be found in the Appendix A). In other words, $\mathcal{L}_\pi$ takes the form of the usual Galileons in 4D plus an extra contribution proportional to the $G_6$ scalar potential and its derivatives. Let us note, however, that the $G_6$ terms can be reabsorbed in the usual 4D Horndeski theory. Indeed, due to the fact that Horndeski theory is the most general scalar-tensor theory of gravity in four-dimensional spacetime with second-order equations of motion, after reduction, we can get rid of all $G_6$ summands and its derivatives by the following redefinition of scalar potentials:
\begin{subequations}
    \begin{flalign}
        &F \rightarrow \hat{F} + \frac{3}{2}{X}^{2} \int \left(G_{6\pi \pi \pi \pi} {X}^{-1}+2G_{6 X \pi\pi\pi\pi}\right)\,\,{\rm d}X - \frac{9}{2}G_{6\pi \pi \pi \pi} {X}^{2}+\frac{3}{2}X \int G_{6 X \pi\pi\pi\pi} X\,\,{\rm d}X,&\\
&K \rightarrow \hat{K} + \frac{9}{2}X \int \left(G_{6\pi \pi \pi} {X}^{-1}+2G_{6 X \pi \pi \pi}\right)\,\,{\rm d}X - \frac{21}{2}X G_{6\pi \pi \pi}+\frac{3}{2}\int X G_{6 X \pi \pi \pi}\,\,{\rm d}X,&\\
&{G_{4}} \rightarrow \hat{G}_{4} + \frac{3}{2}X \int \left(G_{6\pi \pi} {X}^{-1}+2G_{6\pi \pi X}\right)\,\,{\rm d}X-3G_{6\pi \pi} X,&\\
&{G_{5}} \rightarrow \hat{G}_{5} + 3\int \left(G_{6\pi} {X}^{-1}+2G_{6\pi X}\right)\,\,{\rm d}X&
    \end{flalign}\label{eqn redef}
\end{subequations}
thus the Lagrangian $\mathcal{L}_{\pi}$ can be written explicitly as 4D Horndeski with only four independent scalar potentials  $\mathcal{L}_{\pi}(\hat{F},\,\hat{K},\,\hat{G}_4,\,\hat{G}_5)$ (this is more easily checked at the level of the equations of motion). On the other hand, we define $\mathcal{L}_{\phi}$ such that in it we include all contributions to the action without the 4-vector $A^\mu$ and also not contributing to Horndeski theory in 4D, $\, \mathcal{L}_{\pi}(\hat{F},\,\hat{K},\,\hat{G}_4,\,\hat{G}_5)$. Namely, it takes the form,
\begin{subequations}
\begin{eqnarray}
     \mathcal{L}_{\phi} =\mathcal{L}_{K\phi}+\mathcal{L}_{4\phi}+\mathcal{L}_{5\phi}+\mathcal{L}_{6\phi}\,,
\end{eqnarray}
\begin{dmath}
    \mathcal{L}_{K\phi}=\dfrac{1}{\phi} K \phi^{;\alpha} \pi_{;\alpha}\,,
\end{dmath}  
\begin{dmath}
    \mathcal{L}_{4\phi}=\dfrac{2}{\phi}{G_{4}} \left(\Box\phi\right) +\dfrac{4}{\phi}G_{4X} \left(\Box\pi\right) \phi^{;\alpha} \pi_{;\alpha}\,,
\end{dmath}  
\begin{dmath}
    \mathcal{L}_{5\phi}= \dfrac{1}{\phi}{G_{5}} \left(\left(\Box\phi\right) \left(\Box\pi\right) - \phi^{;\alpha\beta} \pi_{;\alpha\beta} \right) - \frac{1}{2\phi}{G_{5}} R \phi^{;\alpha} \pi_{;\alpha} + \dfrac{1}{\phi} G_{5X} \phi^{;\alpha} \pi_{;\alpha} \left({\left(\Box\pi\right)}^{2} - \pi_{;\alpha\beta} \pi^{;\alpha\beta}\right)\,,
\end{dmath}  
\begin{dmath}
    \mathcal{L}_{6\phi} =    \dfrac{6}{\phi}{G_{6}} G^{\alpha\beta} \phi_{;\alpha\beta} +\frac{4}{\phi}G_{6XX}\phi^{;\alpha} \pi_{;\alpha} \left({\left(\Box\pi\right)}^{3} - \left(\Box\pi\right) \pi_{;\alpha\beta} \pi^{;\alpha\beta} + 2 \pi_{;\alpha\beta} \pi^{;\beta}\;_{\gamma} \pi^{;\gamma\alpha}\right) +\dfrac{6}{\phi}G_{6X} \left(\left(\Box\phi\right) {\left(\Box\pi\right)}^{2} - \left(\Box\phi\right) \pi_{;\alpha\beta} \pi^{;\alpha\beta} -2 \left(\Box\pi\right) \phi^{;\alpha\beta} \pi_{;\alpha\beta}\right) +\dfrac{12}{\phi}G_{6X} G^{\alpha\beta} \pi_{;\alpha\beta} \phi^{;\gamma} \pi_{;\gamma}+\dfrac{12}{\phi}G_{6X} \phi^{;\alpha\beta} \pi_{;\beta\gamma} \pi^{;\gamma}\;_{\alpha}\,.
\end{dmath}
\end{subequations}
Finally, we write the action for the 4-vector as
\begin{subequations}
\begin{equation}
    \mathcal{L}_{A} =\mathcal{L}_{4A}+\mathcal{L}_{5A}+\mathcal{L}_{6A}\,,
\end{equation}
\begin{dmath}
    \mathcal{L}_{4A} = -\frac{{\phi}^{2}}{4}{G_{4}} F_{\alpha\beta} F^{\alpha\beta}  + {\phi}^{2}G_{4X} F_{\alpha \gamma} F^{\alpha}\,_{\beta}  \pi^{;\beta} \pi^{;\gamma} \label{eqn L4A}
\end{dmath}

\begin{eqnarray}
    \mathcal{L}_{5A} &=& {G_{5}} \phi^2 \left(\frac{1}{2}F^{\alpha \beta} F_{\alpha}\,^{\gamma} \pi_{;\beta \gamma} - \frac{1}{8}F^{\alpha \beta} F_{\alpha \beta} \Box\pi  +\frac{1}{2}F^{\alpha \beta} \nabla^{\gamma}{F_{\alpha \gamma}} \pi_{;\beta} \right)\nonumber\\
    &+& {G_{5}} \phi^2 \left(\frac{3}{2 \phi}F^{\alpha \beta} F_{\alpha}\,^{\gamma} \phi_{;\beta} \pi_{;\gamma}   -\frac{3}{8 \phi}F^{\alpha \beta} F_{\alpha \beta} \phi^{;\gamma} \pi_{;\gamma} \right)\nonumber\\
  &+& G_{5X} \phi^2 \left(\frac{1}{2}F^{\alpha \beta} F_{\alpha}\,^{\gamma} \left(\Box\pi\right) \pi_{;\beta} \pi_{;\gamma}  - \frac{1}{2}F^{\alpha \beta} F^{\gamma \delta} \pi_{;\alpha \gamma} \pi_{;\beta} \pi_{;\delta}\right)\label{eqn L5A}
\end{eqnarray}

\begin{align}
    \mathcal{L}_{6A}=&{G_{6}} \phi^2 \left(\frac{3}{8}F_{\alpha \beta} F^{\alpha \beta} R - \frac{9}{4 \phi}F_{\alpha \beta} F^{\alpha \beta} \left(\Box\phi\right) +\frac{9 \phi^2}{64}F_{\alpha \beta} F_{\gamma \delta} F^{\alpha \beta} F^{\gamma \delta} +3F_{\alpha \beta} F_{\gamma}\,^{\alpha} R^{\gamma \beta} \right.\nonumber&\\
    &- \frac{9}{2\phi}F_{\alpha \beta} F_{\gamma}\,^{\alpha} \phi^{;\gamma \beta}  - \frac{9\phi^2}{32}F_{\alpha \beta} F_{\gamma}\,^{\alpha} F_{\delta}\,^{\beta} F^{\gamma \delta} - \dfrac{9}{\phi^2}F_{\alpha \beta} F_{\gamma}\,^{\alpha} \phi^{;\beta} \phi^{;\gamma} -\dfrac{9}{\phi}F_{\alpha}\,^{\beta} \nabla^{\gamma}{F_{\beta \gamma}} \phi^{;\alpha}\nonumber&\\
    &+\frac{3}{2}\nabla^{\alpha}{F_{\alpha \beta}} \nabla^{\gamma}{F_{\gamma}\,^{\beta}}+\frac{9}{4}R^{\alpha \gamma \beta \delta} F_{\alpha \beta} F_{\gamma \delta} -\dfrac{3}{\phi} F^{\alpha \beta} \nabla^{\gamma}{F_{\alpha \beta}} \phi_{;\gamma} +\dfrac{3}{\phi} F_{\alpha}\,^{\beta} \nabla^{\alpha}{F_{\beta \gamma}} \phi^{;\gamma}\nonumber&\\
    & \left.- \frac{9}{2\phi^2}F_{\alpha \beta} F^{\alpha \beta} \phi_{;\gamma} \phi^{;\gamma}  - \frac{15}{16}\nabla^{\alpha}{F_{\beta \gamma}} \nabla_{\alpha}{F^{\beta \gamma}}+\frac{3}{8}\nabla^{\alpha}{F_{\beta \gamma}} \nabla^{\beta}{F_{\alpha}\,^{\gamma}}\right)\nonumber&\\
    &+ G_{6X} \left( - \frac{3}{4}F_{\alpha \beta} F^{\alpha \beta} {\left(\Box\pi\right)}^{2} - \frac{9}{2\phi}F_{\alpha \beta} F^{\alpha \beta} \left(\Box\pi\right) \phi^{;\gamma} \pi_{;\gamma} +\frac{3}{2}R F_{\alpha \beta} F_{\gamma}\,^{\alpha} \pi^{;\beta} \pi^{;\gamma}\right.\nonumber\\
    &+\frac{3}{4}F_{\alpha \beta} F^{\alpha \beta} \pi_{;\gamma \delta} \pi^{;\gamma \delta}+\frac{9\phi^2}{8} F_{\alpha \beta} F_{\gamma \delta} F_{f}\,^{\alpha} F^{\gamma \delta} \pi^{;\beta} \pi^{;f} -6F_{\alpha \beta} F_{\gamma}\,^{\alpha} \left(\Box\pi\right) \pi^{;\gamma \beta}\nonumber&\\
    &- \dfrac{9}{\phi}F_{\alpha \beta} F_{\gamma}\,^{\alpha} \pi^{;\gamma \beta} \phi^{;\delta} \pi_{;\delta} -\dfrac{18}{\phi}F_{\alpha \beta} F_{\gamma}\,^{\alpha} \left(\Box\pi\right) \phi^{;\gamma} \pi^{;\beta} -6F_{\alpha}\,^{\beta} \left(\Box\pi\right) \nabla^{\gamma}{F_{\beta \gamma}} \pi^{;\alpha}\nonumber&\\
    &+6F_{\alpha \beta} F_{\gamma}\,^{\alpha} \nabla^{\gamma}{\pi_{;\delta}} \pi^{;\beta \delta}+3F_{\alpha \beta} F_{\gamma \delta} R^{\alpha \gamma} \pi^{;\beta} \pi^{;\delta} - \frac{9 \phi^2 }{4}F_{\alpha \beta} F_{\gamma}\,^{\alpha} F_{\delta}\,^{\beta} F_{f}\,^{\gamma} \pi^{;\delta} \pi^{;f}\nonumber &\\
    &-\dfrac{18}{\phi} F_{\alpha \beta} F_{\gamma \delta} \pi^{;\alpha \gamma} \phi^{;\beta} \pi^{;\delta} +6F_{\alpha \beta} \nabla^{\gamma}{F_{\gamma \delta}} \pi^{;\alpha \delta} \pi^{;\beta} - \frac{9}{2}F_{\alpha \beta} F_{\gamma \delta} \pi^{;\alpha \gamma} \pi^{;\beta \delta}\nonumber&\\
    &\left.+6F_{\alpha}\,^{\beta} \nabla^{\gamma}{F_{\beta \delta}} \pi_{;\gamma}\;^{\delta} \pi^{;\alpha}+\dfrac{18}{\phi}F_{\alpha \beta} F_{\gamma}\,^{\alpha} \pi^{;\gamma \delta} \phi_{;\delta} \pi^{;\beta}\right)\nonumber&\\
    &-3 \phi^2 G_{6XX} \left(F_{\alpha \beta} F_{\gamma}\,^{\alpha} \pi^{;\beta} \pi^{;\gamma} {\left(\Box\pi\right)}^{2}+2F_{\alpha \beta} F_{\gamma \delta} \left(\Box\pi\right) \pi^{;\alpha \gamma} \pi^{;\beta} \pi^{;\delta}\right.\nonumber&\\
    &\left.-F_{\alpha \beta} F_{\gamma}\,^{\alpha} \pi_{;\delta \kappa} \pi^{;\delta \kappa} \pi^{;\beta} \pi^{;\gamma}-2F_{\alpha \beta} F_{\gamma \delta} \pi_{;\alpha}\;^{\kappa} \pi_{;\kappa}\;^{\gamma} \pi^{;\beta} \pi^{;\delta}\right)\label{eqn L6A}&
\end{align}
\end{subequations}\\

 $\mathcal{L}_{4A}$ and the first line of $\mathcal{L}_{5A}$ are the Lagrangians for the photon  such that the graviton of Horndeski theory --- in the case of a frozen Dilaton--- is luminal even with the general scalar potentials $G_4(\pi,\,X)$ and  $G_5(\pi)$ (namely, if $G_{5,X}=0$ and so on), respectively. In the {\it case with a general Dilaton}, a Luminal graviton  is {\it also automatically} obtained if we also consider the second line in $\mathcal{L}_{5A}$. The remaining parts of the Lagrangian contain $G_6(\pi,X)$ and  $G_{5X}(\pi,X)$. As we will see below, in the case that the latter two scalar potentials are present, the graviton and the vector modes propagate in general at {\it different speeds}, unless there is a specific choice of scalar potentials and background solutions. Lagrangians containing "non-luminal" contributions will be generally named $\mathcal{L}_{NL}$. An exception to the latter is a branch of background solutions where luminality of all perturbations but a scalar mode is achieved even with $G_5(\pi,X)$ and $G_6(\pi,X)$. This is shown in section \ref{sec lum branch}.

\section{Notation for the linearized analysis}\label{sec linearization}

We are concerned with IR modifications of GR and of its accompanying photon that could be relevant for dark energy (From now on it will be assumed that we discuss 4D theories, unless otherwise stated). Thus, we will examine the evolution of the FLRW spatially flat, isotropic and homogeneous background, and its stability against first order perturbations of the tensor, vector and scalar modes.

\subsection{Decomposition into irreducible components}
We write the perturbed metric  and 4-vector as
\begin{eqnarray}
    ds^2&=&h_{\mu\nu}\text{d}x^{\mu}\text{d}x^{\nu}=(\eta_{\mu\nu}+\delta h_{\mu\nu})\text{d}x^{\mu}\text{d}x^{\nu}\,,\label{eqn perturbed metric}\\
    A^{\mu}&=&A^{(0)\,\mu}+\delta A^{\mu}\,, \label{eqn perturbed 4vector}
\end{eqnarray}
while the Galileon $\pi(x^{\mu})$ and the Dilaton  $\phi(x^{\mu})$ are splitted as $\pi(t)+\delta \pi(x^{\mu})$ and $\phi(t)+\delta \phi(x^{\mu})$ respectively, and $\pi,\, \phi$ will be understood as background fields or not depending on the context in linearized expressions.

The background, spatially flat FLRW metric is written as usual
\[ds^2=\eta_{\mu\nu}\text{d}x^{\mu}\text{d}x^{\nu} = dt^2 - a(t)^2 \left(dx^2 + dy^2 + dz^2\right)\,, \]
where $a(t)$ is the scale factor. On the other hand, the transverse vector background $A^{(0)\,i}$ is chosen identically to zero by isotropy considerations, and the other components of $A^{(0)\,\mu}$ also vanish on-shell.

The decomposition of perturbations $\delta h_{\mu \nu}, \delta A^{\mu}, \delta \pi$ and $\delta \phi$ into helicity components in the general case has the form
\begin{subequations}
\begin{align}
&\delta h_{00}=2 \Phi \\
&\delta h_{0 i}= - \partial_{i} \beta + Z_i^T, \\
&\delta h_{i j}=-2 \Psi \delta_{i j}-2 \partial_{i}\partial_{j} E - \left(\partial_{i} W_j^T+\partial_{j} W_i^T\right)+h^T_{i j},\\
&\delta A_{0} = \gamma,\\
&\delta A_{i} = \partial_{i}\alpha + A_{i},\\
&\delta\pi = \chi, \\
&\delta\phi = \varphi,
\end{align}    \label{eqn decomposition}
\end{subequations}
where $\Phi, \beta, \Psi, E, \chi, \varphi, \alpha, \gamma$ are scalar fields, $Z_{i}^{T}, W_{i}^{T}, A_{i}$ are transverse two-component vector fields ($\partial_{i}Z_{i}^{T} = \partial_{i} W_{i}^{T} = \partial_{i} A_{i} = 0$) and $h^T_{ij}$ is a  transverse traceless two-component tensor. 

The evolution of the cosmological background is determined by three equations $\mathcal{E}_{f}=\partial \mathcal{L}/\partial f = 0$ where $f$ is one of the following: $\pi,\, \phi,\, h_{00},\, h_{ij}$. There is a redundancy due to little gauge invariance, and thus, we can express, for instance, $\mathcal{E}_{\pi}$ in terms of the others and their time derivatives. By looking at the equations of the background fields and the linearized expressions, it becomes evident that it is useful to define $H_\phi = \dot{\phi}/\phi$ in analogy with our Hubble parameter $H = \dot{a}/a$.

\subsection{From 5D diffeomorphisms, to 4D little gauge and $U(1)$ gauge invariant vector}

The initial 5D theory in Eqn. (\ref{lagrangian}) is invariant under small coordinate transformations in $M_5$
\[\label{diff} x^M \rightarrow x^M - \xi^M.\]
As usual, upon dimensional reduction the invariance under 5D diffeomorphisms turns into invariance under 4D diffeomophisms and $U(1)$ gauge transformations of the vector field $A_\mu$. All fields under the small coordinate shift~\eqref{diff} are transformed as follows:
\begin{subequations}
\begin{eqnarray}
      h_{\mu \nu} &\rightarrow& h_{\mu \nu} + \partial_{\alpha}\eta_{\mu \nu} \xi^{\alpha} + \partial_{\mu}{\xi^{\alpha}} \eta_{\nu \alpha} + \partial_{\nu}{\xi^{\alpha}} \eta_{\mu \alpha},\label{eqn 4Ddiff}\\
    A_{\mu} &\rightarrow& A_\mu + \xi^{\alpha}\partial_{\alpha}{A^{(0)}_{\mu}} + A_{\alpha}^{(0)} \partial_{\mu}{\xi^{\alpha}} + \partial_{\mu}{\xi^{4}},\\
    \varphi &\rightarrow& \varphi + \xi^{\alpha} \partial_{\alpha}\phi\,,\\
    \chi &\rightarrow& \chi + \xi^{\alpha} \partial_{\alpha}\pi,
\end{eqnarray}
\end{subequations}
where $\xi^{\mu} = \left(\xi_0, \xi^i_T + \delta^{i j} \partial_{j} \xi_{S}\right)^{\text{T}}$.

\section{Luminal Gravitational Waves with $\mathcal{L}_{4A}$ and $\mathcal{L}_{5A}$ {\it vs.} non-luminality with   $\mathcal{L}_{NL}$ }\label{sec luminality}

\subsection{The graviton}

Expanding to second order the action (\ref{lagrangian_after_reduction}) with (\ref{eqn perturbed metric})-(\ref{eqn decomposition}), and using the equations for the background fields and its derivatives to vanish the $h_{ij}^2$ term, we obtain the quadratic action for the tensor perturbation of the metric
\[\label{tensor_action} S_{tensor}^{(2)}=\int \mathrm{d} t \mathrm{~d}^3 x a^3 \phi \left[\mathcal{G}_\tau \left(\dot{h}_{i j}\right) ^2-\frac{\mathcal{F}_\tau}{a^2}\left( \overrightarrow{\nabla} h_{i j}\right)^2\right]\,,\]
where $\mathcal{G}_\tau$ and $\mathcal{F}_\tau$ are functions of the background fields $\pi(t),\, \phi(t), \, a$. They are given in the Appendix \ref{secapp coefficients}. They take the same form as in the usual Horndeski theory, up to $H_\phi=\frac{\dot{\phi}}{\phi}$ contributions.

Let us recall that the squared speed of the tensor modes 
\begin{equation}
    c_g^2=\frac{\mathcal{F}_\tau}{\mathcal{G}_\tau}\,,
\end{equation}
is different from unity even in the Horndeski theory case, namely, even if $H_\phi=0$. For instance, in the case $H_\phi=0$ and with zero scalar potential $G_5$, the speed takes the well known\footnote{Let us recall that if $H_\phi=0$ the scalar potential $G_6$ can be fully reabsorbed by (\ref{eqn redef}).} form with $\mathcal{F}_\tau=2G_4$ and $\mathcal{G}_\tau=2 (G_4-2\dot{\pi}^2\,G_{4X})$. Thus, only if $G_4$ is a function of $\pi$ -but not of $X$- can the speed of the graviton be exactly unity. In other words, because for a minimally coupled photon the speed of light is constant ($c_A^2=1$ in our units), Horndeski theories with the minimal Photon are, in general, heavily constrained by the precise experimental bounds on the speed of gravitational waves and light  \cite{Ezquiaga:2017ekz},
\begin{equation}
    \vert \frac{c_g}{c_A}-1 \vert \leq 5 \times 10^{-16}\,.
\end{equation}
A way around this strict constraint is to consider Horndeski theory complemented by the Photon-Scalar couplings shown in the actions (\ref{eqn L4A}) and (\ref{eqn L5A}) for the general scalar potentials $G_4(\pi,X)$ and $G_5(\pi)$. In those cases, even for nonzero $H_\phi$, the speed of the Photon changes with time in exactly the same way as the speed of graviton so that  $\frac{c_g(t)}{c_A(t)}=1$ is preserved throughout cosmic evolution. The key aspect is that the effective metric of both the graviton and the Photon are modified by the same Gravito-electric Galileon. Indeed, let us now compute the perturbations to the cosmological background in the vector sector.

\subsection{Vector sector and the Photon}
Noticing that the combination $$V_{i} = \dot{W}_{i} - 2 H W_{i} + Z_{i}$$ is gauge-invariant by the transformations 
\begin{subequations}
\begin{flalign}
&Z_{i} \rightarrow  Z_{i} - {a}^{2} \dot{\xi^{i}_T}, \\
&W_{i} \rightarrow W_{i} + a^2 \xi^i_T,
\end{flalign}  
\end{subequations}
obtained from (\ref{eqn 4Ddiff}), and  expanding to second order the action (\ref{lagrangian_after_reduction}) with (\ref{eqn perturbed metric})-(\ref{eqn decomposition}), the vector sector can be written as
\begin{flalign}
  \delta S^2_{vector} = \int &\mathrm{d}t\,\mathrm{d}^3x\,a^3 \phi  \left(\dfrac{{\phi}^{2}}{a^2} \left(\mathcal{G}_{V}\: \left(\dot{\vec{A}}\right)^2\right)- \frac{{\phi}^{2}}{a^2} \mathcal{F}_{V}\: \left( \dfrac{\overrightarrow{\nabla}}{a}\times \vec{A}\right)^2 + \mathcal{K} \: \left( \dfrac{\overrightarrow{\nabla}}{a}\times \vec{V}\right)^2 \right)\,,\label{eqn vector sector}
\end{flalign}
where $\mathcal{G}_{V},\, \mathcal{F}_{V},\, \mathcal{K} $ are functions of the background fields. They are given in the Appendix \ref{secapp coefficients}. 

From (\ref{eqn vector sector}) it is clear that the vector perturbations of the metric $\vec{V}$ and the vector perturbations of the 4-vector $A^{\mu}$ decouple. The former is nondynamical, and thus we can write the final action for the vector modes as 
\begin{flalign}
  \delta S^2_{vector} = \int &\mathrm{d}t\,\mathrm{d}^3x\,a  \phi^3  \left( \mathcal{G}_{V}\: \left(\dot{\vec{A}}\right)^2-  \mathcal{F}_{V}\: \left( \dfrac{\overrightarrow{\nabla}}{a}\times \vec{A}\right)^2 \right)\,,\label{eqn vector sector final}
\end{flalign}
such that they propagate with the speed squared
\begin{equation}
    c_A^2=\frac{\mathcal{F}_{V}}{\mathcal{G}_{V}}\,.
\end{equation}
As anticipated from the discussion in the previous section, for the theories (\ref{lagrangian_after_reduction}) with only $F(\pi,X),\,K(\pi,X),\,G_4(\pi,X)$ and $G_5(\pi)$ general scalar potentials, we can note that
\begin{equation}
\dfrac{\mathcal{F}_{V}}{\mathcal{G}_{V}}=\dfrac{\mathcal{F}_{\tau}}{\mathcal{G}_{\tau}}
\end{equation}
and thus, the speed of the graviton and the 4-vector are the same throughout cosmic evolution $\frac{c_g}{c_A}=1$ \cite{Mironov:2024idn}. Now, provided the experimental bounds that set the speed of GWs very close to the speed of light, it is reasonable -at least in these cases- to identify the 4-vector with the Photon relevant for cosmic evolution. We discuss this identification from other perspective in section \ref{sec screening}. We will generally denote these as theories with an exactly "luminal graviton" or Luminal extensions of Horndeski theory.

It is worth to note that a recent work \cite{Babichev:2024kfo} found that, at least in a cosmological background, the $G_4(\pi,X)$ Luminal extensions of Horndeski theory -namely, with nonminimal Photon-Scalar couplings $\mathcal{L}_{4A}$ and with $G_5\equiv 0$- can be reproduced by disformal transformations of already known luminal {\it Beyond Horndeski theories} with a minimally coupled photon. Thus, in this sense and as far as it is known, only on a cosmological background, the $\mathcal{L}_{4A}$ Luminal extension of Horndeski maps to a Beyond Horndeski theory.

\subsection{The 4-vector and the Photon with $\mathcal{L}_{4A}$ and $\mathcal{L}_{5A}$}\label{sec screening}

The common practice in IR modifications of GR, and in particular, in Horndeski theory, is to identify the tensor modes of perturbations about cosmological background with the graviton of GR. This identification is clearly motivated in first place, by the two propagated polarizations and in second place, by the fact that the modifications of GR due to the Horndeski scalar are screened in solar system tests. See for instance \cite{ Vainshtein:1972sx,Babichev:2013usa,Koyama:2013paa,Kobayashi:2019hrl}. The latter is specially important provided the stringent experimental bounds, otherwise, there would be little point in naming them "IR modifications of {\it gravity}" (the theory valid in the solar system) in the first place.

Similarly, with regards to the Vector-Galileon couplings discussed before $\mathcal{L}_{4A},\,\mathcal{L}_{5A}$, there is as a matter of fact, an experimental motivation to try to identify that $U(1)$ gauge invariant vector coupled to the Horndeski scalar with the Photon: the strict bounds on the similarity between the speeds of gravitational waves and light by the event GW170817 \cite{LIGOScientific:2017vwq}. These suggest that a graviton even in modified gravity theories {\it may} propagate at the same speed as light. Assuming the latter, $\mathcal{L}_{4A},\,\mathcal{L}_{5A}$ together with Horndeski theory (in the case of a constant dilaton) expand the theory space of allowed\footnote{See however \cite{Fernandes:2022zrq} for an alternative.} theories in what respects these speed tests, at least within the Horndeski class\footnote{Namely, while keeping second order equations of motion. See however  \cite{Babichev:2024kfo} for a recent discussion in Beyond Horndeski theory.}. Now, how to match the IR scalar modification of the familiar Photon of Maxwell electrodynamics to laboratory tests is an important challenge. This was briefly discussed in connection with the Vainshtein screening of GR modifications in \cite{Mironov:2024idn}. And a related discussion on experimental bounds for these sort of scalar-Photon couplings was advanced in \cite{Babichev:2024kfo}. Although this question merits more investigation, it falls way beyond the scope of this work.

\section{Scalar modes, a luminal branch of solutions  and the stability analysis of the FLRW background}\label{sec stability}

\subsection{Scalar sector}
Expanding to second order the action (\ref{lagrangian_after_reduction}) with (\ref{eqn perturbed metric})-(\ref{eqn decomposition}), and partly using the gauge freedom to set the longitudinal component $E$ to vanish,  the quadratic action for the scalar perturbations has the form
\begin{flalign}
  \delta S^{2}_{scalar} = \int &\mathrm{d}t\,\mathrm{d}^3x\,a^3 \phi  \left({A_{1}}\: {\left(\dot{\Psi}\right)}^{2}+{A_{2}}\: \dfrac{(\overrightarrow{\nabla}\Psi)^2}{a^2}+{A_{3}} \: {\Phi}^{2}+{A_{4}}\: \Phi \dfrac{\overrightarrow{\nabla}^2\beta}{a^2}+{A_{5}}\: \dot{\Psi} \dfrac{\overrightarrow{\nabla}^2\beta}{a^2} +{A_{6}}\: \Phi \dot{\Psi} \right.\nonumber \\
    +{A_{7}}& \: \Phi \dfrac{\overrightarrow{\nabla}^2{\Psi}}{a^2}+{A_{8}} \: \Phi \dfrac{\overrightarrow{\nabla}^2{\chi}}{a^2}+{A_{9}}\: \dfrac{\overrightarrow{\nabla}^2{\beta}}{a^2} \dot{\chi} +{A_{10}} \:\chi\ddot{\Psi}+{A_{11}} \:\Phi \dot{\chi}+{A_{12}}\: \chi \dfrac{\overrightarrow{\nabla}^2\beta}{a^2}+{A_{13}} \:\chi \dfrac{\overrightarrow{\nabla}^2\Psi}{a^2} \nonumber\\
    +{A_{14}}&\: {\left(\dot{\chi}\right)}^{2}+{A_{15}}\: \dfrac{(\overrightarrow{\nabla}\chi)^2}{a^2}  +{A_{17}}\: \Phi \chi +{A_{18}}\: \chi \dot{\Psi}+{A_{19}}\: \Psi \chi + {A_{20}} \:{\chi}^{2}+{C_{1}} \: \dot{\Psi} \dot{\varphi} +{C_{2}} \: \dot{\Psi} \varphi+{C_{4}}\: \chi \ddot{\varphi}\nonumber\\
     + {C_{5}}&\: \chi \dot{\varphi}+{C_{7}} \:\dot{\varphi} \dfrac{\overrightarrow{\nabla}^2{\beta}}{a^2}+{C_{8}}\: \varphi \dfrac{\overrightarrow{\nabla}^2{\beta}}{a^2}+ {C_{9}}\:\Phi \dot{\varphi} + {C_{10}}\: \varphi \dfrac{\overrightarrow{\nabla}^2{\chi}}{a^2}+{C_{11}}\: \varphi \dfrac{\overrightarrow{\nabla}^2{\Psi}}{a^2}+{C_{12}}\: \varphi \dfrac{\overrightarrow{\nabla}^2{\Phi}}{a^2} \nonumber\\
     + {\phi}^2&\left.\left( {B_{1}}\:  \dfrac{{(\overrightarrow{\nabla}{\dot{\alpha}})}^{2}}{a^2}+ {B_{2}}\: \gamma \dfrac{\overrightarrow{\nabla}^2{\alpha}}{a^2}-2 B_1 \dot{\gamma} \dfrac{\overrightarrow{\nabla}^2{\alpha}}{a^2} +{B_{1}} \: \dfrac{(\overrightarrow{\nabla}{\gamma})^{2}}{a^2}\right)\right),\label{eqn scalar sector}
\end{flalign}
where coefficients $A_i, B_i, C_i$ are  combinations of the Lagrangian functions and their derivatives, written in terms of the background fields. The explicit form of these coefficients can be found in the Appendix \ref{secapp coefficients}.

A first important aspect is that the scalar perturbations of $A^\mu$, namely $\alpha,\, \gamma$ do not mix with the scalar perturbations of metric, Galileon or Dilaton. We will use this fact later on to discuss the stability of the scalar modes.

 The equations of motion obtained from the variations on $\beta, \Phi, \gamma$, and $\alpha$ are constraints. The equations from the variations on $\gamma$ and $\alpha$ are equivalent. They can be written as,
\begin{subequations}
\label{constraints}
  \begin{flalign}
  \delta\Phi:\quad &\dfrac{\overrightarrow{\nabla}^2{\beta}}{a^2}  = -\frac{1}{A_4}\Bigg(A_{17} \chi+A_8 \dfrac{\overrightarrow{\nabla}^2 \chi}{a^2} +A_{11} \dot{\chi} +2 A_3 \Phi + C_{12}\dfrac{\overrightarrow{\nabla}^2 \varphi}{a^2} + C_9 \dot{\varphi}+A_7 \dfrac{\overrightarrow{\nabla}^2 \psi}{a^2} +A_6 \dot{\psi} \Bigg) \\
  \delta\beta:\quad &\Phi = -\frac{1}{ A_4}\left(A_{12} \chi+A_9\dot{\chi}+C_8\varphi +C_7 \dot{\varphi}+A_5\dot{\psi}\right),\\
  \delta\gamma:\quad & \dfrac{\overrightarrow{\nabla}^2{\gamma}}{a^2} = \dfrac{\overrightarrow{\nabla}^2{\dot{\alpha}}}{a^2}.
\end{flalign}
\end{subequations}

%\begin{subequations}
%\label{constraints}
 % \begin{flalign}
  %\delta\Phi:\quad &\dfrac{\overrightarrow{\nabla}^2{\beta}}{a^2}  = \frac{1}{3 A_4^2}\Bigg(-\left(2{A_{1}} \dfrac{\overrightarrow{\nabla}^2{\Psi}}{a^2}-9{A_{4}} \dot{\Psi}+{C_{1}} \dfrac{\overrightarrow{\nabla}^2{\varphi}}{a^2}+3{C_{9}} \dot{\varphi}\right) {A_{4}}\nonumber\\
 %  &- 2 A_3 \left(2{A_{1}}  \dot{\Psi} + C_2 \varphi + C_1 \dot{\varphi}\right)\Bigg) \\
 % \delta\beta:\quad &\Phi = \frac{1}{3 A_4}\left(2{A_{1}} \dot{\Psi}+ C_2 \varphi + C_1 \dot{\varphi}\right),\\
 % \delta\gamma:\quad & \dfrac{\overrightarrow{\nabla}^2{\gamma}}{a^2} = \dfrac{\overrightarrow{\nabla}^2{\dot{\alpha}}}{a^2}.
%\end{flalign}
%\end{subequations}
The $\gamma$ constraint turns out to be trivial due to the relation $\frac{1}{2}{B_{2}} +\dot{{B_{1}}} +B_1 \left(3H_{\phi} +H \right) = 0$.

Now let us use the residual gauge freedom and set $\chi = 0$. Using the constraints \eqref{constraints} back in the quadratic Lagrangian (\ref{eqn scalar sector}), we integrate out $\Phi, \beta, \gamma$. The terms containing $\alpha$ are canceled automatically. The resulting action contains two dynamic variables, $\Psi$ and $\varphi$:
\begin{eqnarray}
      \delta S^{2}_{scalar} = \int \mathrm{d}t\,\mathrm{d}^3x\,a^3 \phi \Big(&&{\left(\dot{\Psi}\right)}^{2}  \mathcal{G}_{S}^{\Psi} +\mathcal{G}_{S}^{\Psi \varphi} \dot{\Psi} \dot{\varphi}  + {\left(\dot{\varphi}\right)}^{2}  \mathcal{G}_{S}^{\varphi}  - \mathcal{F}_{S}^{\Psi}\dfrac{\left(\overrightarrow{\nabla} \Psi\right)^2}{a^2} - \mathcal{F}_{S}^{\Psi\varphi} \dfrac{\overrightarrow{\nabla} \Psi}{a} \dfrac{\overrightarrow{\nabla} \varphi}{a}  -  \mathcal{F}_{S}^{\varphi} \dfrac{\left(\overrightarrow{\nabla} \varphi \right)^2}{a^2}\nonumber\\
  &&+\mathcal{R} \dot{\Psi} \varphi   + M_{\varphi} {\varphi}^{2} \Big),
\end{eqnarray}
where
\begin{subequations}
\begin{equation}
    \begin{array}{ccc}
      \mathcal{G}_{S}^{\Psi} = -{A_{1}}+\dfrac{4}{9}\dfrac{ {A_{3}} {{A_{1}}}^{2}}{{A_{4}^2}} \,,  & \mathcal{G}_{S}^{\Psi \varphi} = \dfrac{2}{9}\dfrac{\left(2{A_{3}} {C_{1}}+3{A_{4}} {C_{9}}\right) {A_{1}}}{{A_4}^2} \,, & \mathcal{G}_{S}^{\varphi} = \dfrac{1}{9} \dfrac{{A_{3}} {{C_{1}}}^{2}}{A_{4}^2}+\dfrac{1}{3} \dfrac{{C_{1}} {C_{9}}}{A_4}\,,
    \end{array}
\end{equation}

\begin{equation}
    \begin{array}{ccc}
      \mathcal{F}_{S}^{\Psi} = -{A_{2}}- \dfrac{1}{a \phi} \dfrac{d}{dt}\left[\dfrac{2 {{A_{1}}}^{2}\phi a}{9A_{4}}\right]  \,,& \mathcal{F}_{S}^{\Psi\varphi} =  C_{11} -  \dfrac{1}{a^2}  \dfrac{d}{dt}\left[{\dfrac{2A_1 C_1 a^2}{9 A_4}}\right] \,, & \mathcal{F}_{S}^{\varphi} = - \dfrac{\phi}{a^3} \dfrac{d}{dt}\left[\dfrac{{a}^{3} {{C_{1}}}^{2}}{18 \phi A_4} \right]\,,
    \end{array}
\end{equation}

\begin{equation}
    \begin{array}{cc}
   \mathcal{R} =  \dfrac{4}{9}\dfrac{{A_{1}} {A_{3}} {C_{2}}}{{A_{4}}^2}\,, & M_{\varphi} =  -\dfrac{1}{9 a^4 }\dfrac{d}{dt} \left[\dfrac{{A_{3}}   {{C_{1}}} C_2 {a}^{4}}{A_4^2}   \right]-\dfrac{1}{6 \phi a^3}\dfrac{d}{dt}\left[\dfrac{{C_{2}} {C_{9}} \phi  {a}^{3}}{A_4} \right]\,,  
    \end{array}
\end{equation}
\end{subequations}
Defining a kinetic matrix $\mathcal{G}_S$ and a "gradient" matrix $\mathcal{F}_S$,
\begin{equation}
\begin{array}{cc}
   \mathcal{G}_S= \left( \begin{array}{cc}
      \mathcal{G}_{S}^{\Psi}    & \frac{1}{2} \mathcal{G}_{S}^{\Psi\varphi}  \\
        \frac{1}{2} \mathcal{G}_{S}^{\Psi\varphi} & \mathcal{G}_{S}^{\varphi}
    \end{array}\right)\,, & \mathcal{F}_S= \left( \begin{array}{cc}
      \mathcal{F}_{S}^{\Psi}    & \frac{1}{2} \mathcal{F}_{S}^{\Psi\varphi}  \\
        \frac{1}{2} \mathcal{F}_{S}^{\Psi\varphi} & \mathcal{F}_{S}^{\varphi}
    \end{array}\right)\,,
    \end{array}
\end{equation}
it is easily verified that $\mathcal{G}_S$ has non-zero determinant, and thus there are two degrees of freedom,
\begin{equation}
    \mathrm{det}\, \mathcal{G}_{S}  =  - \frac{1}{9} \dfrac{A_{1}}{{A_{4}^2}} \left({A_{1}} {{C_{9}}}^{2}+{A_{3}} {{C_{1}}}^{2}+3{A_{4}} {C_{1}} {C_{9}}\right) \,.
\end{equation}
Furthermore, a field rotation that diagonalizes the kinetic matrix does not diagonalize $\mathcal{F}_S$. Thus, the speeds of propagation of the two scalar modes can be found as eigenvalues of the matrix ${\mathcal{G}_{S}}^{-1} \mathcal{F}_{S}$. They are not factorized and their expressions are cumbersome, but we can express their sum and product in terms of the determinant of both matrices using the property that the trace of a matrix is the sum of its eigenvalues, and the determinant of a matrix is the product of eigenvalues:
\begin{subequations}
\begin{eqnarray}
    &(c_{S}^{(1)})^2 \, (c_{S}^{(2)})^2 =  \dfrac{\mathrm{det} \mathcal{F}_{S}}{\mathrm{det} \mathcal{G}_{S}},&\\
    &(c_{S}^{(1)})^2 + (c_{S}^{(2)})^2 = \dfrac{1}{\mathrm{det} \mathcal{G}_{S}} (\mathcal{G}^{\Psi}_{S} \mathcal{F}^{\varphi}_{S} - \frac{1}{2}\mathcal{G}^{\Psi\varphi}_{S} \mathcal{F}^{\Psi\varphi}_{S}+\mathcal{G}^{\varphi}_{S} \mathcal{F}^{\Psi}_{S} )\,,&
\end{eqnarray}
\end{subequations}
where,
\begin{eqnarray}
    \mathrm{det}\,\mathcal{F}_{S} =  \frac{\phi}{18 a^3}\left({A_{2}} + \frac{2}{9 a \phi } \dfrac{d}{dt} \left[\dfrac{\phi a {{A_{1}}}^{2}}{{A_{4}}}\right]\right) \dfrac{d}{dt} \left[\dfrac{{{C_{1}}}^{2}  {a}^{3}}{{{A_{4}}}{\phi}}\right] - \frac{1}{4}{\left(- {C_{11}} + \frac{2}{9 a^2} \dfrac{d}{dt}\left[\dfrac{{a}^{2} {A_{1}} {C_{1}}}{{A_{4}}}\right]\right)}^{2} .
\end{eqnarray}
A particular branch of background solutions where we solve explicitly the speeds of the scalars is given below, in section \ref{sec lum branch}.

\subsection{Stability of a nonsingular cosmological background against perturbations}

Let us now discuss the stability of the FLRW background.  Let us first assume a definite sign for the Dilaton background $\phi(t)$. This is a natural assumption because otherwise at the vanishing of $\phi(t)$ the 5D metric would be degenerate. Indeed, let us recall that the components of the  4-vector background are also vanishing on-shell and by isotropy.

Now, requiring that the tensor and vector modes are not ghosts we demand $\mathcal{G}_\tau>0$ and $\mathcal{G}_V>0$. Furthermore, to avoid gradient instabilities we require that  $\mathcal{F}_\tau>0$ and  $\mathcal{F}_V>0$. Finally, we assume conventional asymptotics in that $\mathcal{F}_\tau$ has a nonzero lower bound in the asymptotic past and future (See for instance  \cite{Kobayashi:2016xpl,Ageeva:2021yik}). Let us note in passing that for the Luminal extensions of Horndeski theory - namely, in the cases with $\mathcal{L}_{4A}$ and $\mathcal{L}_{5A}$ but without $G_6$ or $X$ dependance on $G_5$- only two conditions are really required. Indeed, in that case the no-ghost and stability of the graviton obviously imply the same for the Photon.

On the other hand, in the scalar sector it must be required that $ \mathrm{det}\,\mathcal{G}_{S}>0$ and $ \mathrm{det}\,\mathcal{F}_{S}>0$.

However, let us now show that well-known No-Go theorems \cite{Libanov:2016kfc, Kobayashi:2016xpl} for all-time stable, {\it nonsingular} cosmological solutions also apply in this case, even with the general Galileon and Dilaton. Thus even if local stability can be achieved at some time for some model, it is certain that -assuming conventional asymptotics and not special cases (See  \cite{Kobayashi:2016xpl,Ageeva:2021yik,Mironov:2022quk})- a gradient instability will arise earlier or later in the evolution.\\

To this end let us recall that by the analysis of the previous section, the scalar perturbations of $A^\mu$ decouple from the scalar perturbations of the metric, Galileon and Dilaton. In fact, the former are non-dynamical. Thus the quadratic action for the scalar sector obtained from (\ref{lagrangian_after_reduction})  is equivalent on-shell to a quadratic action obtained by setting $A^\mu$ to zero, or $\mathcal{L}_A\rightarrow 0$ in (\ref{lagrangian_after_reduction}). Namely, we can obtain the same scalar dynamics considering only the $\mathcal{L}_\pi$ and $\mathcal{L}_\phi$ in (\ref{lagrangian_after_reduction}). Now, let us recall that using the redefinition of scalar potentials (\ref{eqn redef}) in $\mathcal{L}_\pi$, we can rewrite this part of the action only as a function of $\hat{F},\,\hat{K},\, \hat{G}_4,\,\hat{G}_5$ (namely, without $G_6$). On the other hand, upon the redefinition (\ref{eqn redef}) we obtain  $\mathcal{L}_\phi=\mathcal{L}_\phi(\hat{F},\,\hat{K},\, \hat{G}_4,\,\hat{G}_5,\,G_6) $. Now, following \cite{Padilla:2012dx, Akama:2017jsa} let us define the following multi-scalar potentials 
\begin{subequations}
\begin{eqnarray}
  &G_2 = \hat{F}(\pi, X) + \dfrac{1}{\phi} \hat{K}(\pi, X) \phi^{;\alpha} \pi_{;\alpha},&\\
  &G_3^{\pi} = \hat{K}(\pi, X) + \dfrac{4}{\phi} \hat{G}_{4 X} \phi^{;\alpha} \pi_{;\alpha},&\\
  &G_{3}^{\phi} = \dfrac{2}{\phi} \hat{G}_{4}(\pi, X),\\
  &\tilde{G}_4 = \hat{G}_{4} + \dfrac{1}{2 \phi} \hat{G}_{5}(\pi,X) \phi^{;\alpha} \pi_{;\alpha},&\\
  &G_5^{\pi} = \hat{G}_{5}(\pi, X) + \dfrac{9}{\phi} G_{6X} \phi^{;\alpha} \pi_{;\alpha},&\\
  &G_5^{\phi} = \dfrac{6}{\phi} G_{6},&
\end{eqnarray}
\end{subequations}
Thus we can rewrite $\mathcal{L}_\pi+\mathcal{L}_\phi$ as
\begin{flalign}
  \begin{aligned}
\mathcal{L}= & G_2-G_{3}^{I} \square \phi^I+G_4 R+G_{4,\langle I J\rangle}\left(\square \phi^I \square \phi^J-\nabla_\mu \nabla_\nu \phi^I \nabla^\mu \nabla^\nu \phi^J\right) \\
& +G_{5}^{I} G^{\mu \nu} \nabla_\mu \nabla_\nu \phi^I-\frac{1}{6} G_{5,(J K\rangle}^{I}\left[\square \phi^I \square \phi^J \square \phi^K-3 \square \phi^{(I} \nabla_\mu \nabla_\nu \phi^J \nabla^\mu \nabla^\nu \phi^{K)}\right. \\
& \left.+2 \nabla_\mu \nabla_\nu \phi^I \nabla^\nu \nabla^\lambda \phi^J \nabla_\lambda \nabla^\mu \phi^K\right],
\end{aligned}
\end{flalign}
where, in our case, $\phi^I$ are the fields $\pi$ and $\phi$, $X^{I J}:=h^{\mu \nu} \partial_\mu \phi^I \partial_\nu \phi^J$, and the index $_{<IJ>}$ denotes the symmetrized derivative $f_{,\langle I J\rangle}:= \left(\partial f / \partial X^{I J}+\partial f / \partial X^{J I}\right) / 2$. Thus, the No-Go theorem proved in \cite{Akama:2017jsa} applies to the theories discussed in this note.

\subsection{A "Luminal" branch of background solutions}\label{sec lum branch}

At this point it is relevant to discuss a cross-check of all the computations. This discussion will reveal an interesting aspect of the action (\ref{lagrangian_after_reduction}) when considered within the general context of vector-scalar Galileons, or SVT theories.

Since the 4D metric, the 4-vector and the Dilaton are different components of the same 5D metric, it would be expected that, for instance, their speeds are the same. As we saw before, this happens in many cases, but not all. In particular, the speed of the 4-vector and the graviton is different if the scalar potential $G_5$ depends on $X$ or $G_6$ is present. The reason behind this difference is that we have compactified one dimension and broken isotropy in the spatial dimensions. As {\it a cross-check of our results}, let us restore isotropy and homogeneity in the spacelike dimensions in 5D by setting $\phi(t)=a(t)$ (in fact, the discussion below only requires us to assume $H_\phi=H$), in other words, by considering FLRW in 5D. In that case another of the equations for the background fields becomes redundant --- namely, the equation for the Dilaton background $\phi$ ($\mathcal{E}_{\phi}=0$) --- and thus, as expected, only $\mathcal{E}_{h_{00}}=0,\, \mathcal{E}_{h_{11}}=0$ are independent. Furthermore,  we find in that case that both {\it the speeds of the graviton and the 4-vector coincide} even including the general potentials $G_5(\pi,X),\, G_6(\pi,X)$. Finally, the action coefficients are simplified as
\[C_1 = A_1, \quad C_2 = C_8=0,\quad C_9 = - A_4, \quad C_{11} = - 2 A_2,\]
and the speeds for the scalar modes can be explicitly written as 
\begin{subequations}
    \begin{eqnarray}
          &(c_S^{(1)})^2 = {A_4}^2 \dfrac{\dfrac{1}{a^2} \dfrac{d}{dt}\left[\dfrac{{a}^{2} {{A_{1}}}^{2}}{{A_{4}}}\right] + 6 A_2}{4{A_{1}} {{A_{4}}}^{2} - 2{{A_{1}}}^{2} {A_{3}}} ,&\\[2ex]
        &(c_S^{(2)})^2 = -3\dfrac{A_{2}}{{A_{1}}} = \dfrac{\mathcal{F}_\tau}{\mathcal{G}_\tau}.&
    \end{eqnarray}
\end{subequations}
Thus, we verify that the speed of one of the scalars --- the extra scalar mode due to the Dilaton ---, $(c_S^{(2)})^2$, coincides with the speed of the tensor and vector  modes for all scalar potentials $G_5(\pi,X)$ and $G_6(\pi,X)$. 

Let us stress, however, that this special case {\it lacks physical significance} within the context of the KK dimensional reduction. Indeed, for the latter it was natural to assume that the radius of the compactified fifth dimension is small, thus necessarily breaking 5D spatial isotropy and homogeneity, and such that there is in principle no reason to assume that $H_\phi=H$.

On the other hand, let us discuss the later result more broadly: Within the context of all higher derivative SVT, the last result simply says that the theory (\ref{lagrangian_after_reduction}) is a subclass with a "Luminal" branch of background solutions. Namely, without discussing its connection to higher dimensions, consider the action (\ref{lagrangian_after_reduction}) as an SVT theory for the metric, two higher derivative scalars $\phi$ and $\pi$ --- {\bf here regarded on equal footing} --- and for a $U(1)$ gauge invariant higher derivative vector $A^\mu$, such that the equations of motion are of second order. This theory has the defining property that it has a particular {\it branch of cosmological background solutions $(\phi(t)=a(t))$ on which Luminality of all perturbations, with the exception of a scalar mode, is achieved for all scalar potentials up to $G_6(\pi,X
)$}. Despite its unphysical character in connection with the higher dimensional origin, this case could be, however, relevant on the search for Vector-Galileon couplings from a broader perspective, beyond our specific method of KK reductions.

\section{Conclusions}\label{sec conclusions}

We expanded on the results shown in a recent letter \cite{Mironov:2024idn} and showed a broad class of $U(1)$ gauge invariant vector/ multi-scalar Galileons in 4D: namely, a theory with higher derivatives of a $U(1)$ gauge invariant vector and two scalars in the action, but with second order equations of motion for all of the fields. It was obtained from the dimensional reduction of Galileons in five dimensions ($5D$). We did not aim for completeness, nor for a complete classification of $U(1)$ gauge invariant vector Galileons. Nevertheless, the theories shown in this work provide a consistent way to  couple vectors to IR scalar modifications of gravity --- without inducing ghosts, and keeping gauge invariance---  in the aim to explore more universal couplings of the scalar of dark energy to other matter. Furthermore, they are a specially relevant class of vector Galileons because first, a subclass of these theories leads to a luminal graviton in 4D even with general $G_4(\pi,X)$ and $G_5(\pi)$ scalar potentials, second, for the whole class of theories up to $G_6(\pi,X)$ there is a {\it particular  branch} of background solutions where most modes (tensor, vector and one scalar) propagate at the same speed, and third, because being connected to higher dimensional theories they could be simpler to analyze in some cases, for instance, in the look for Black-Hole solutions in 4D.

\section*{Acknowledgements}
The work on this project
has been supported by Russian Science Foundation grant № 24-72-10110,

\href{https://rscf.ru/project/24-72-10110/}{  https://rscf.ru/project/24-72-10110/}.

\newpage
\section{Appendices}
\subsection{Details for the dimensional reduction}\label{sec decomp}
In this Appendix we collect the expressions for all five-dimensional scalar structures in terms of four-dimensional ones:
\begin{flalign}
  &\widehat{R} = R-\frac{2}{\phi} \Box{\phi} + \frac{1}{4}F_{\alpha \beta} F^{\alpha \beta} {\phi}^{2},&
\end{flalign}
\begin{flalign}
  &\hat{\Box}\pi = \Box \pi + \frac{1}{\phi}\phi^{;\alpha} \pi_{;\alpha},&
  \end{flalign}
\begin{flalign}
&\pi_{;MN}\pi^{;MN} = \pi_{;\mu\nu}\pi^{;\mu\nu} -\frac{1}{2}F_{\alpha \beta} F^{\alpha \gamma} \pi^{;\beta} \pi_{;\gamma} {\phi}^{2} +\frac{1}{\phi^2}\left(\phi^{;\alpha} \pi_{;\alpha}\right)^2,&
\end{flalign}
\begin{flalign}
&\pi_{;MN}\pi^{;MP}\pi_{;P}^{\;\;N} = \pi_{;\mu\nu}\pi^{;\mu \kappa}\pi_{;\kappa}\;^{\nu}-\frac{3 {\phi}^{2} }{4}F^{\alpha \beta} F^{\gamma \delta} \pi_{;\alpha} \pi_{;\gamma} \pi_{;\beta\delta} &\nonumber\\
&-\frac{3 \phi }{4}F^{\alpha \beta} F_{\alpha \delta}  \pi_{;\beta} \pi^{;\delta}\phi_{;\gamma}\pi^{;\gamma} +\frac{1}{\phi^3}\left(\phi^{;\alpha} \pi_{;\alpha}\right)^3,&
\end{flalign}
\begin{flalign}
&\hat{G}^{MN}\pi_{;MN} = G^{\alpha \beta} \pi_{;\alpha \beta} +\frac{1}{\phi}\Box\phi \Box\pi-\frac{1}{\phi}\phi^{;\alpha \beta}\pi_{;\alpha \beta} - \frac{\phi^2}{8}F_{\alpha \beta} F^{\alpha \beta} \Box\pi  +\frac{\phi^2}{2}F^{\alpha \beta} F_{\alpha \gamma} \pi_{;\beta}\;^{\gamma} &\nonumber\\
&+ \frac{3 \phi }{2}F^{\alpha \beta} F_{\alpha \delta} \phi_{;\beta}\pi^{;\delta} + \frac{\phi^2}{2}F^{\alpha \beta} \nabla^{\gamma}{F_{\alpha \gamma}} \pi_{;\beta}-\frac{3\phi}{8}F_{\alpha \beta} F^{\alpha \beta} \pi^{;\gamma}\phi_{;\gamma} - \frac{1}{2\phi} {R} \phi^{;\alpha} \pi_{;\alpha},
\end{flalign}
\begin{flalign}
&\pi^{;A B} \pi_{;B C} \pi^{;C D} \pi_{;A D} = \pi^{;\alpha \beta} \pi_{;\beta \gamma} \pi^{;\gamma \delta} \pi_{;\delta\alpha} - {\phi}^{2} F^{\alpha \beta} F^{\gamma \delta} \pi_{;\alpha}^{\;\;\epsilon} \pi_{;\gamma \epsilon} \pi_{;\beta} \pi_{;\delta} +\frac{{\phi}^{4}}{8}  \left(F^{\alpha \beta} F_{\alpha \delta} \pi_{;\beta} \pi^{;\delta}\right)^2  &\nonumber\\
&- \phi F^{\alpha \beta} F^{\gamma \delta} \pi_{;\alpha \gamma}  \pi_{;\beta}\pi_{;\delta} \phi^{;\epsilon} \pi_{;\epsilon}  - F^{\alpha \beta} F_{\alpha \gamma} \pi^{;\gamma} \left(\phi_{;\delta} \pi^{;\delta}\right)^2  + \frac{1}{\phi^4} \left(\phi_{;\alpha} \pi^{;\alpha}\right)^4, &
\end{flalign}

\begin{flalign}
       &\hat{R}_{MN}\hat{R}^{MN} = R^{\alpha \beta} R_{\alpha \beta}-\dfrac{2}{\phi} R^{\alpha \beta} \phi_{;\alpha\beta} +F^{\alpha \beta} F_{\alpha}\,^{\gamma} R_{\beta \gamma} {\phi}^{2}+ \dfrac{1}{\phi^2} \phi^{;\alpha\beta} \phi_{;\alpha\beta} &\nonumber\\
    &- \phi F^{\alpha \beta} F_{\alpha}\,^{\gamma} \phi_{;\beta\gamma}+\frac{{\phi}^{4}}{4}F^{\alpha \beta} F_{\alpha}\,^{\gamma} F_{\beta}\,^{\delta} F_{\gamma \delta} -\frac{9}{2}F^{\alpha \beta} F_{\alpha}\,^{\gamma} \phi_{;\beta} \phi_{;\gamma} -3\phi F^{\alpha \beta} \nabla^{\gamma}{F_{\alpha \gamma}} \phi_{;\beta} &\\
    &+ \frac{{\phi}^{2}}{2}\nabla^{\alpha}{F_{\alpha}\,^{\beta}} \nabla^{\gamma}{F_{\beta \gamma}} +\dfrac{1}{\phi^2} \left(\Box\phi\right)^2  + \frac{\phi}{2}F^{\alpha \beta} F_{\alpha \beta} \Box\phi+\frac{{\phi}^{4}}{16}F^{\alpha \beta} F_{\alpha \beta} F^{\gamma \delta} F_{\gamma \delta}, &\nonumber
\end{flalign}
\begin{flalign}
    &\hat{R}^{ABCD} \hat{R}_{ABCD} = R^{\alpha \beta \gamma \delta} R_{\alpha \beta \gamma \delta}+3{\phi}^{2} F^{\alpha \beta} F^{\gamma \delta} R_{\alpha \gamma \beta \delta}+\frac{3}{8}F^{\alpha \beta} F_{\alpha \beta} F^{\gamma \delta} F_{\gamma \delta} {\phi}^{4}&\nonumber\\
    &+\frac{5}{8}F^{\alpha \beta} F_{\alpha}\,^{\gamma} F_{\beta}\,^{\delta} F_{\gamma \delta} {\phi}^{4} -\frac{5}{4}\nabla^{\gamma}{F^{\alpha \beta}} \nabla_{\gamma}{F_{\alpha \beta}} {\phi}^{2}- 4 \phi F^{\alpha \beta} \phi_{;\gamma} \left(\nabla^{\gamma}{F_{\alpha \beta}} - \nabla_{\alpha}{F_{\beta}\,^{\gamma}}\right) &\nonumber\\
    &-6 \left(F^{\alpha \beta} \phi_{;\gamma}\right)^2-6F^{\alpha \beta} F_{\alpha}\,^{\gamma} \phi_{;\beta}\phi_{;\gamma} + \frac{\phi^2}{2}\nabla^{\gamma}{F^{\alpha \beta}} \nabla_{\alpha}{F_{\gamma \beta}} +\dfrac{4}{\phi^2} \phi^{;\alpha\beta}\phi_{;\alpha\beta}+2 \phi F^{\alpha \beta} F_{\alpha}\,^{\gamma} \phi_{;\beta\gamma}, & 
\end{flalign}
\begin{flalign}
    &\hat{R}^{AB} \pi_{;AB} = - \dfrac{1}{\phi}\phi^{;\alpha\beta}\pi_{;\alpha\beta} -\dfrac{1}{\phi^2}\left(\Box\phi\right) \phi^{;\alpha}\pi_{;\alpha} +R^{\alpha\beta} \pi_{;\alpha\beta} + \frac{{\phi}^{2}}{2}F^{\alpha\beta} F_{\alpha}\,^{\gamma} \pi_{;\beta\gamma} & \nonumber\\
    & + \frac{{\phi}^{2}}{2}F^{\alpha\beta} \nabla^{\gamma}{F_{\alpha \gamma}}\pi_{;\beta}  + \frac{3 \phi}{2}F^{\alpha \beta} F_{\alpha}\,^{\gamma} \phi_{;\beta} \pi_{;\gamma}-\frac{\phi}{4}F^{\alpha\beta} F_{\alpha\beta} \phi^{;\gamma}\pi_{;\gamma},&
\end{flalign}
\begin{flalign}
   &\hat{R}^{A B} \pi_{;A}^{\;\;\;C} \pi_{;B C} = R^{\alpha\beta} \pi_{;\alpha}\;^{\gamma} \pi_{;\beta \gamma}-\dfrac{1}{\phi}\phi^{;\alpha\beta} \pi_{;\alpha}\;^{\gamma} \pi_{;\beta\gamma}-\dfrac{1}{\phi^3} \left(\phi^{;\alpha} \pi_{;\alpha}\right)^2\left(\Box\phi\right)  &\nonumber\\
    &+F^{\alpha\beta} F_{\alpha}\,^{\gamma} \left(\frac{{\phi}^{2}}{2} \pi_{;\beta}\;^{\delta} \pi_{;\gamma\delta}  - \frac{{\phi}^{4}}{8} F_{\beta}\,^{\delta} F_{\gamma}\,^{\kappa} \pi_{;\delta} \pi_{;\kappa}  +\frac{3}{2} \phi_{;\beta} \phi^{;\delta} \pi_{;\gamma} \pi_{;\delta} +\frac{\phi}{4} \left(\Box\phi\right) \pi_{;\beta} \pi_{;\gamma} \right)&\nonumber\\
    &+F^{\alpha\beta} F^{\gamma\delta} \left( -\frac{{\phi}^{2}}{4} R_{\alpha\gamma} \pi_{;\beta} \pi_{;\delta} + \frac{\phi}{4} \phi_{;\alpha \gamma} \pi_{;\beta} \pi_{;\delta}  + \frac{3\phi}{2} \pi_{;\alpha\gamma} \phi_{;\beta} \pi_{;\delta} +\frac{{\phi}^{4}}{16} F_{\alpha\beta} F_{\gamma}\,^{\kappa} \pi_{;\delta} \pi_{;\kappa}  \right)&\nonumber\\
    & -\frac{ {\phi}^{2}}{2}F^{\alpha\beta} \nabla^{\gamma}{F_{\gamma}\,^{\delta}} \pi_{;\alpha\delta} \pi_{;\beta} + \frac{1}{2}F^{\alpha\beta} \nabla^{\gamma}{F_{\alpha \gamma}} \phi^{;\delta} \pi_{;\beta} \pi_{;\delta} \phi + \frac{1}{4}\left(F^{\alpha\beta} \phi^{;\gamma} \pi_{;\gamma}\right)^2,&
\end{flalign}

\begin{flalign}
 & \hat{R}^{A B C D} \pi_{;A C} \pi_{;B D} = R^{\alpha\beta\gamma\delta} \pi_{;\alpha\gamma} \pi_{;\beta \delta}-\dfrac{2}{\phi^2}\phi^{;\alpha\beta}\pi_{;\alpha\beta} \phi^{;\gamma} \pi_{;\gamma} +F^{\alpha\beta} \left( \frac{3  {\phi}^{2}}{4}F^{\gamma\delta} \pi_{;\alpha\gamma} \pi_{;\beta\delta}\right.&\nonumber\\
 &\left.+ {\phi}^{2}\nabla^{\gamma}{F_{\alpha}\,^{\delta}} \pi_{;\beta} \pi_{;\gamma\delta}+3\phi F_{\alpha}\,^{\gamma} \phi^{;\delta} \pi_{;\beta} \pi_{;\gamma\delta} -\frac{\phi}{2}F_{\alpha}\,^{\gamma} \phi^{;\delta} \pi_{;\delta} \pi_{;\beta \gamma} \phi - \frac{{\phi}^{4}}{8}F_{\alpha}\,^{\gamma} F_{\beta}\,^{\delta} F_{\gamma}\,^{\kappa} \pi_{;\delta} \pi_{;\kappa} \right)&\nonumber\\
 &-\frac{1}{2}F^{\alpha\beta} F^{\gamma\delta} \phi_{;\alpha\gamma} \pi_{;\beta} \pi_{;\delta} \phi.&
\end{flalign}

\subsection{Dynamics of the cosmological background}\label{secapp background equations}
The equations of motion for the background fields, from the Lagrangian \eqref{lagrangian_after_reduction} have the form:
\begin{flalign}
  \delta h^{0 0}: \quad &F =  2F_{X} {\dot{\pi}}^{2} -K_{\pi} {\dot{\pi}}^{2} + 2 K_{X} {\dot{\pi}}^{3} \left(H_{\phi}+3H\right)- 6 {G_{4}} H\left(H_{\phi}+{H}\right) &\nonumber\\
  &+ 24 G_{4X} {\dot{\pi}}^{2} H \left({H}+H_{\phi}\right) + 24 G_{4XX} H {\dot{\pi}}^{4} \left(H_{\phi}+{H}\right) - 2 G_{4\pi} \dot{\pi} \left(H_{\phi}+3H\right) &\nonumber\\
  &-4G_{4X \pi} {\dot{\pi}}^{3} \left(H_{\phi}+3H\right) +10G_{5X} H^2 {\dot{\pi}}^{3} \left(3H_{\phi} +{H}\right)+ 4 G_{5XX} H^2 {\dot{\pi}}^{5} \left(3H_{\phi} +{H}\right)&\\
  & -9 G_{5\pi} H {\dot{\pi}}^{2} \left({H} +  H_{\phi}\right) -6 G_{5X \pi} H {\dot{\pi}}^{4} \left(H_{\phi}+{H}\right)-18{G_{6}} H_{\phi} {H}^{3}&\nonumber\\
  &+108G_{6X} {\dot{\pi}}^{2} H_{\phi} {H}^{3} -18 G_{6\pi} H^2 \dot{\pi} \left(3H_{\phi}+{H}\right) -72 G_{6X \pi} H^2 {\dot{\pi}}^{3} \left(3 H_{\phi} + {H}\right)&\nonumber\\
  & +216G_{6XX} {\dot{\pi}}^{4} H_{\phi} {H}^{3} + 48G_{6XXX} H_{\phi} {H}^{3} {\dot{\pi}}^{6}-24G_{6 X X \pi } H^2 {\dot{\pi}}^{5} \left(3H_{\phi} + {H}\right),\nonumber
\end{flalign}
\begin{flalign}
  \delta h^{i i}: \quad &0 = F-2K_{X} \ddot{\pi} {\dot{\pi}}^{2}- K_{\pi} {\dot{\pi}}^{2} +2{G_{4}} \left(2\dot{H}+3{H}^{2}+2H H_{\phi}+\dot{H_{\phi}}+{H_{\phi}}^{2}\right)&\nonumber\\
  &-4G_{4X} {\dot{\pi}}^{2} \left(3{H}^{2}+2H H_{\phi}+2\dot{H}+\dot{H_{\phi}}+{H_{\phi}}^{2}\right)-4G_{4X} \dot{\pi} \ddot{\pi} \left(2H+H_{\phi}\right)&\nonumber\\
  &+2 G_{4\pi} \left(\ddot{\pi}+\dot{\pi} \left(H_{\phi}+2H\right)\right)+2G_{4\pi\pi}{\dot{\pi}}^{2}+4G_{4\pi X} \ddot{\pi} {\dot{\pi}}^{2} -4G_{4\pi X} {\dot{\pi}}^{3} \left(H_{\phi}+2H\right)&\nonumber\\
  &-8G_{4XX} \ddot{\pi} {\dot{\pi}}^{3} \left(H_{\phi}+2H\right)+G_{5 \pi} {\dot{\pi}}^{2} \left(3{H}^{2}+\dot{H_{\phi}}+{H_{\phi}}^{2}+2\dot{H}+2H H_{\phi}\right)&\nonumber\\
  &+2G_{5 \pi} \dot{\pi} \ddot{\pi} \left(H_{\phi}+2H\right)-4G_{5X} {\dot{\pi}}^{3} \left(H \dot{H}+{H}^{3}+2H_{\phi} {H}^{2}+H_{\phi} \dot{H}+H \dot{H_{\phi}}+H {H_{\phi}}^{2}\right)&\nonumber\\
  &-6G_{5X} \ddot{\pi} H {\dot{\pi}}^{2} \left(2 H_{\phi}+{H}\right)-2G_{5\pi X} H {\dot{\pi}}^{4} \left({H}+2 H_{\phi}\right)+2G_{5\pi X} \ddot{\pi} {\dot{\pi}}^{3} \left(H_{\phi}+2H\right)&\nonumber\\
  &-4G_{5XX} H \ddot{\pi} {\dot{\pi}}^{4} \left({H}+2 H_{\phi}\right)+G_{5 \pi \pi} {\dot{\pi}}^{3} \left(H_{\phi}+2H\right)+24G_{6\pi X X} H \ddot{\pi} {\dot{\pi}}^{4} \left(2 H_{\phi}+H\right)&\nonumber\\
  &+ 6 {G_{6}} H \left(\dot{H_{\phi}} {H}+{H} {H_{\phi}}^{2}+2 H_{\phi} \dot{H}+2H_{\phi} {H}^{2}\right)+6 G_{6\pi} H \ddot{\pi} \left(2 H_{\phi}+{H}\right)&\nonumber\\
  &+6G_{6\pi} \dot{\pi} \left(2H \dot{H_{\phi}}+2H {H_{\phi}}^{2}+2H \dot{H}+2{H}^{3}+5H_{\phi} {H}^{2}+2H_{\phi} \dot{H}\right)+6 G_{6\pi \pi} H {\dot{\pi}}^{2} \left(2 H_{\phi}+{H}\right)&\nonumber\\
  &-36G_{6X} \dot{\pi} \ddot{\pi} H_{\phi} {H}^{2}-24G_{6X} H {\dot{\pi}}^{2} \left(2 H_{\phi} \dot{H}+2H_{\phi} {H}^{2}+\dot{H_{\phi}} {H}+{H} {H_{\phi}}^{2}\right)&\nonumber\\
  &+48G_{6\pi X} H \ddot{\pi} {\dot{\pi}}^{2} \left(2 H_{\phi}+{H}\right)+24G_{6\pi X} {\dot{\pi}}^{3} \left(H_{\phi} \dot{H}+H_{\phi} {H}^{2}+H \dot{H_{\phi}}+H {H_{\phi}}^{2}+H \dot{H}+{H}^{3}\right)&\nonumber\\
  &-144G_{6XX} \ddot{\pi} {\dot{\pi}}^{3} H_{\phi} {H}^{2}+12 G_{6\pi \pi X} H {\dot{\pi}}^{4} \left(2 H_{\phi}+ {H}\right)-24G_{6\pi X X} {\dot{\pi}}^{5} H_{\phi} {H}^{2}&\nonumber\\
  &-24G_{6XX} {\dot{\pi}}^{4} \left(2H H_{\phi} \dot{H}+2H_{\phi} {H}^{3}+\dot{H_{\phi}} {H}^{2}+{H}^{2} {H_{\phi}}^{2}\right)-48\ddot{\pi} {\dot{\pi}}^{5} G_{6XXX} H_{\phi} {H}^{2},&
\end{flalign}
\begin{flalign}
  \delta \phi: \quad &0 = {G_{4}} \left(-2H H_{\phi}+\dot{H}-\dot{H_{\phi}}-{H_{\phi}}^{2}+3{H}^{2}\right)+ G_{4\pi} \dot{\pi} \left(H-H_{\phi}\right)+4G_{4XX} \ddot{\pi} {\dot{\pi}}^{3} \left(H_{\phi}-H\right)&\nonumber\\
  &+2G_{4X} {\dot{\pi}}^{2} \left(-3{H}^{2}+2H H_{\phi}-\dot{H}+\dot{H_{\phi}}+{H_{\phi}}^{2}\right)-2G_{4X} \dot{\pi} \ddot{\pi} \left(H-H_{\phi}\right)+2 G_{4\pi X} {\dot{\pi}}^{3} \left(H_{\phi}-H\right)&\nonumber\\
  &+\dfrac{1}{2} G_{5 \pi} {\dot{\pi}}^{2} \left(3{H}^{2}-2H H_{\phi} - \dot{H_{\phi}} - {H_{\phi}}^{2}+\dot{H}\right) +G_{5 \pi} \dot{\pi} \ddot{\pi} \left(H-H_{\phi}\right)- \frac{1}{2}G_{5 \pi \pi}{\dot{\pi}}^{3} \left(  H_{\phi}- H\right)&\nonumber\\
  &+6G_{5X} H \ddot{\pi} {\dot{\pi}}^{2} \left(H_{\phi}-{H}\right)+2 G_{5X} {\dot{\pi}}^{3} \left(2H_{\phi} {H}^{2}-3{H}^{3}-2H \dot{H}+H_{\phi} \dot{H}+H \dot{H_{\phi}}+H {H_{\phi}}^{2}\right)&\nonumber\\
  &+G_{5\pi X} {\dot{\pi}}^{3} \left(H_{\phi}-H\right) \left(2H \dot{\pi}  - \ddot{\pi}\right)-4G_{5XX} H \ddot{\pi} {\dot{\pi}}^{4} \left({H}-H_{\phi}\right)&\nonumber\\
  &+3{G_{6}} H \left(-2H_{\phi} {H}^{2}-\dot{H_{\phi}} {H}-{H} {H_{\phi}}^{2}-2 H_{\phi} \dot{H}+3\dot{H} {H}+3{H}^{3}\right)+18G_{6X} \dot{\pi} H^2 \ddot{\pi} \left(H_{\phi}-{H}\right)&\nonumber\\
   &+12 G_{6X} H {\dot{\pi}}^{2} \left(2H_{\phi} {H}^{2}+2H_{\phi} \dot{H}+\dot{H_{\phi}} {H}+{H} {H_{\phi}}^{2}-3\dot{H} {H}-3{H}^{3}\right)-48 G_{6\pi X} H\ddot{\pi} {\dot{\pi}}^{2} \left( H_{\phi}-{H}\right)&\nonumber\\
   &+3 G_{6\pi} \dot{\pi} \left(-5H_{\phi} {H}^{2}+7{H}^{3}-2H \dot{H_{\phi}}-2H {H_{\phi}}^{2}+4H \dot{H}-2H_{\phi} \dot{H}\right)-6 G_{6\pi} H \ddot{\pi} \left(H_{\phi}-{H}\right)&\nonumber\\
   &-6 G_{6\pi \pi} H {\dot{\pi}}^{2}  \left( H_{\phi}-{H}\right) +12 G_{6XX} H {\dot{\pi}}^{4} \left(2H_{\phi} {H}^{2}+2H_{\phi} \dot{H}+\dot{H_{\phi}} {H}+{H} {H_{\phi}}^{2}-3\dot{H} {H}-36{H}^{3}\right)&\nonumber\\
  &+12G_{6\pi X} {\dot{\pi}}^{3} \left(-H_{\phi} {H}^{2}+2{H}^{3}-H_{\phi} \dot{H}-H \dot{H_{\phi}}-H {H_{\phi}}^{2}+2H \dot{H}\right)+72 G_{6XX}H^2 \ddot{\pi} {\dot{\pi}}^{3} \left(H_{\phi} -{H}\right)&\nonumber\\
  &+12 G_{6\pi X X} H {\dot{\pi}}^{4} \left(H_{\phi}-{H}\right)\left(H \dot{\pi} - 2 \ddot{\pi}\right)-12G_{6\pi \pi X} H{\dot{\pi}}^{4} \left( H_{\phi}- {H}\right)+24G_{6XXX} H^2\ddot{\pi} {\dot{\pi}}^{5} \left(H_{\phi}-{H}\right),&
\end{flalign}
where we denote $H = \dot{a}/a$ and $H_\phi = \dot{\phi}/\phi$, dot denotes the derivative with respect to the cosmic time $t$.

\subsection{Coefficients of the quadratic action}\label{secapp coefficients}
In the tensor sector we defined $\mathcal{G}_\tau=\frac{A_5}{2}$ and $\mathcal{F}_\tau=A_2$, and in the vector sector 
\begin{flalign}
  \mathcal{G}_{V} = B_1,\quad \mathcal{F}_{V} = \dfrac{1}{8}C_{11}\quad \mathcal{K} =12 A_1\,.
\end{flalign}
These coefficients and those shown in the quadratic  action for the scalar sector are,
\begin{flalign}
  &A_1 = -6{G_{4}}+12G_{4X} {\left(\dot{\pi}\right)}^{2}-3G_{5\pi} {\left(\dot{\pi}\right)}^{2}+6G_{5X} {\left(\dot{\pi}\right)}^{3}\left(H+H_{\phi}\right) +72G_{6XX} H H_{\phi} {\left(\dot{\pi}\right)}^{4}&\nonumber\\
  &+72G_{6X} H H_{\phi} {\left(\dot{\pi}\right)}^{2}-36G_{6X \pi} {\left(\dot{\pi}\right)}^{3}\left(H_{\phi}+H\right) -18{G_{6}} H H_{\phi}-18 G_{6\pi} \dot{\pi}\left(H+H_{\phi}\right), 
\end{flalign}
\begin{flalign}
  &{A_{2}} = 2{G_{4}}-G_{5\pi} {\left(\dot{\pi}\right)}^{2}-2G_{5X} \left(\ddot{\pi}\right) {\left(\dot{\pi}\right)}^{2}-12G_{6X} {H_{\phi}}^{2} {\left(\dot{\pi}\right)}^{2}+12G_{6X \pi} \left(\ddot{\pi}\right) {\left(\dot{\pi}\right)}^{2}&\nonumber\\
  &-12G_{6X \pi} H_{\phi} {\left(\dot{\pi}\right)}^{3}-24G_{6XX} H_{\phi} \left(\ddot{\pi}\right) {\left(\dot{\pi}\right)}^{3}-12G_{6X} H_{\phi} \dot{\pi} \left(\ddot{\pi}\right)-12G_{6X} \dot{H_{\phi}} {\left(\dot{\pi}\right)}^{2}&\\
  &+6{G_{6}} \dot{H_{\phi}}+6{G_{6}} {H_{\phi}}^{2}+6G_{6\pi\pi} {\left(\dot{\pi}\right)}^{2}+6G_{6\pi} H_{\phi} \dot{\pi}+6G_{6\pi} \left(\ddot{\pi}\right),&\nonumber
\end{flalign}
\begin{flalign}
  &A_3 = F_{X} {\left(\dot{\pi}\right)}^{2}+4 K_{X} {\left(\dot{\pi}\right)}^{3} \left(H_{\phi}+3H\right) -K_{\pi} {\left(\dot{\pi}\right)}^{2}-2G_{4\pi} \dot{\pi}\left(H_{\phi}+3H\right) +42G_{4X} {H}^{2} {\left(\dot{\pi}\right)}^{2}&\nonumber\\
  &+42G_{4X} H H_{\phi} {\left(\dot{\pi}\right)}^{2}+96G_{4XX} H H_{\phi} {\left(\dot{\pi}\right)}^{4}+96G_{4XX} {H}^{2} {\left(\dot{\pi}\right)}^{4}-10G_{4X \pi} {\left(\dot{\pi}\right)}^{3}\left(H_{\phi}+3H\right)&\nonumber\\
  & -6{G_{4}} H H_{\phi}-6{G_{4}} {H}^{2}-27G_{5X \pi} H H_{\phi} {\left(\dot{\pi}\right)}^{4}-27G_{5X \pi} {H}^{2} {\left(\dot{\pi}\right)}^{4}+78G_{5XX} H_{\phi} {H}^{2} {\left(\dot{\pi}\right)}^{5}&\nonumber\\
  &+26G_{5XX} {H}^{3} {\left(\dot{\pi}\right)}^{5}-18G_{5\pi} {H}^{2} {\left(\dot{\pi}\right)}^{2}-18G_{5\pi} H H_{\phi} {\left(\dot{\pi}\right)}^{2}+90G_{5X} H_{\phi} {H}^{2} {\left(\dot{\pi}\right)}^{3}+30G_{5X} {H}^{3} {\left(\dot{\pi}\right)}^{3}&\nonumber\\
  &+456G_{6XXX} H_{\phi} {H}^{3} {\left(\dot{\pi}\right)}^{6}-504G_{6 X X \pi } H_{\phi} {H}^{2} {\left(\dot{\pi}\right)}^{5}-168G_{6 X X \pi } {H}^{3} {\left(\dot{\pi}\right)}^{5}+972G_{6XX} H_{\phi} {H}^{3} {\left(\dot{\pi}\right)}^{4}&\\
  &-702G_{6X \pi} H_{\phi} {H}^{2} {\left(\dot{\pi}\right)}^{3}-234G_{6X \pi} {H}^{3} {\left(\dot{\pi}\right)}^{3}+306G_{6X} H_{\phi} {H}^{3} {\left(\dot{\pi}\right)}^{2}-36{G_{6}} H_{\phi} {H}^{3}-108G_{6\pi} H_{\phi} \dot{\pi} {H}^{2}&\nonumber\\
  &-36G_{6\pi} \dot{\pi} {H}^{3}+2F_{X X} {\left(\dot{\pi}\right)}^{4}-K_{X \pi} {\left(\dot{\pi}\right)}^{4}+2K_{X X} {\left(\dot{\pi}\right)}^{5}\left(H_{\phi}+3H\right) +24G_{4XXX} H H_{\phi} {\left(\dot{\pi}\right)}^{6}&\nonumber\\
  &+24G_{4XXX} {H}^{2} {\left(\dot{\pi}\right)}^{6}-4G_{4 X X \pi } {\left(\dot{\pi}\right)}^{5}\left(H_{\phi}+3H\right) -6G_{5 X X \pi } H H_{\phi} {\left(\dot{\pi}\right)}^{6}-6G_{5 X X \pi } {H}^{2} {\left(\dot{\pi}\right)}^{6}&\nonumber\\
  &+12G_{5XXX} H_{\phi} {H}^{2} {\left(\dot{\pi}\right)}^{7}+4G_{5XXX} {H}^{3} {\left(\dot{\pi}\right)}^{7}+48G_{6 XXXX} H_{\phi} {H}^{3} {\left(\dot{\pi}\right)}^{8}-72G_{6 XXX \pi} H_{\phi} {H}^{2} {\left(\dot{\pi}\right)}^{7}&\nonumber\\
  &-24G_{6 XXX \pi} {H}^{3} {\left(\dot{\pi}\right)}^{7},&\nonumber
\end{flalign}
\begin{flalign}
  &A_4 = 2K_{X} {\left(\dot{\pi}\right)}^{3}-2{G_{4}}\left(2H+H_{\phi}\right) +8G_{4X} {\left(\dot{\pi}\right)}^{2}\left(2H+H_{\phi}\right) -4G_{4X \pi} {\left(\dot{\pi}\right)}^{3}-2G_{4\pi} \dot{\pi}\nonumber&\\
  &+8G_{4XX} {\left(\dot{\pi}\right)}^{4}\left(H_{\phi}+2H\right) +10G_{5X} {H}^{2} {\left(\dot{\pi}\right)}^{3}+20G_{5X} H H_{\phi} {\left(\dot{\pi}\right)}^{3}-3G_{5\pi} {\left(\dot{\pi}\right)}^{2}\left(2H+H_{\phi}\right)\nonumber&\\
  & -2G_{5X \pi} {\left(\dot{\pi}\right)}^{4}\left(2H+H_{\phi}\right) +8G_{5XX} H H_{\phi} {\left(\dot{\pi}\right)}^{5}+4G_{5XX} {H}^{2} {\left(\dot{\pi}\right)}^{5}+216G_{6XX} H_{\phi} {H}^{2} {\left(\dot{\pi}\right)}^{4}&\\
  &+48G_{6XXX} H_{\phi} {H}^{2} {\left(\dot{\pi}\right)}^{6}-48G_{6 X X \pi } H H_{\phi} {\left(\dot{\pi}\right)}^{5}-24G_{6 X X \pi } {H}^{2} {\left(\dot{\pi}\right)}^{5}+108G_{6X} H_{\phi} {H}^{2} {\left(\dot{\pi}\right)}^{2}\nonumber&\\
  &-144G_{6X \pi} H H_{\phi} {\left(\dot{\pi}\right)}^{3}-72G_{6X \pi} {H}^{2} {\left(\dot{\pi}\right)}^{3}%
-18{G_{6}} H_{\phi} {H}^{2}-36G_{6\pi} H H_{\phi} \dot{\pi}-18G_{6\pi} \dot{\pi} {H}^{2},&\nonumber
\end{flalign}
\begin{flalign}
  &A_5 = -\dfrac{2}{3}A_1,&
\end{flalign}
\begin{flalign}
  &A_6 = -3 A_4,&
\end{flalign}
\begin{flalign}
 & A_7 = \dfrac{2}{3}A_1,&
\end{flalign}
\begin{flalign}
  &A_8 = 2K_{X} {\left(\dot{\pi}\right)}^{2}-2G_{4\pi}+4 G_{4X} \dot{\pi}\left(2H+H_{\phi}\right)+8G_{4XX} {\left(\dot{\pi}\right)}^{3}\left(H_{\phi}+2H\right) -4G_{4X \pi} {\left(\dot{\pi}\right)}^{2}&\nonumber\\
  &+12G_{5X} H H_{\phi} {\left(\dot{\pi}\right)}^{2}+6G_{5X} {H}^{2} {\left(\dot{\pi}\right)}^{2}-2G_{5\pi} \dot{\pi}\left(2H+H_{\phi}\right) -2G_{5X \pi} {\left(\dot{\pi}\right)}^{3}\left(H_{\phi}+2H\right)&\nonumber\\
  & +8G_{5XX} H H_{\phi} {\left(\dot{\pi}\right)}^{4}+4G_{5XX} {H}^{2} {\left(\dot{\pi}\right)}^{4}+48G_{6XXX} H_{\phi} {H}^{2} {\left(\dot{\pi}\right)}^{5}-48G_{6 X X \pi } H H_{\phi} {\left(\dot{\pi}\right)}^{4}&\\
  &-24G_{6 X X \pi } {H}^{2} {\left(\dot{\pi}\right)}^{4}+36G_{6X} H_{\phi} \dot{\pi} {H}^{2}+144G_{6XX} H_{\phi} {H}^{2} {\left(\dot{\pi}\right)}^{3}-48G_{6X \pi} {H}^{2} {\left(\dot{\pi}\right)}^{2}&\nonumber\\
  &-96G_{6X \pi} H H_{\phi} {\left(\dot{\pi}\right)}^{2}-12G_{6\pi} H H_{\phi}%
-6G_{6\pi} {H}^{2},&\nonumber
\end{flalign}
\begin{flalign}
  &A_9 = - A_8,&
\end{flalign}
\begin{flalign}
  &A_{10} = -3 A_8,&
\end{flalign}
\begin{flalign}
  &A_{11} = -2F_{X} \dot{\pi}-4F_{X X} {\left(\dot{\pi}\right)}^{3}+2K_{X \pi} {\left(\dot{\pi}\right)}^{3}+2K_{\pi} \dot{\pi}-4K_{X X} {\left(\dot{\pi}\right)}^{4}\left(H_{\phi}+3H\right) &\nonumber\\
  &-6K_{X} {\left(\dot{\pi}\right)}^{2}\left(H_{\phi}+3H\right) -36G_{4X} H H_{\phi} \dot{\pi}-36G_{4X} \dot{\pi} {H}^{2}-144G_{4XX} H H_{\phi} {\left(\dot{\pi}\right)}^{3}&\nonumber\\
  &-144G_{4XX} {H}^{2} {\left(\dot{\pi}\right)}^{3}+16 G_{4X \pi} {\left(\dot{\pi}\right)}^{2}\left(H_{\phi}+3H\right) +2 G_{4\pi} \left(H_{\phi}+3H\right) -48G_{4XXX} H H_{\phi} {\left(\dot{\pi}\right)}^{5}&\nonumber\\
  &-48G_{4XXX} {H}^{2} {\left(\dot{\pi}\right)}^{5}+8G_{4 X X \pi } {\left(\dot{\pi}\right)}^{4}\left(H_{\phi}+3H\right) -120G_{5XX} H_{\phi} {H}^{2} {\left(\dot{\pi}\right)}^{4}-40G_{5XX} {H}^{3} {\left(\dot{\pi}\right)}^{4}&\nonumber\\
  &-90G_{5X} H_{\phi} {H}^{2} {\left(\dot{\pi}\right)}^{2}-30G_{5X} {H}^{3} {\left(\dot{\pi}\right)}^{2}%
+42G_{5X \pi} {H}^{2} {\left(\dot{\pi}\right)}^{3}+18G_{5\pi} \dot{\pi} {H}^{2}+42G_{5X \pi} H H_{\phi} {\left(\dot{\pi}\right)}^{3}&\nonumber\\
&+18G_{5\pi} H H_{\phi} \dot{\pi}+12G_{5 X X \pi } H H_{\phi} {\left(\dot{\pi}\right)}^{5}+12G_{5 X X \pi } {H}^{2} {\left(\dot{\pi}\right)}^{5}-24G_{5XXX} H_{\phi} {H}^{2} {\left(\dot{\pi}\right)}^{6}&\\
&-8G_{5XXX} {H}^{3} {\left(\dot{\pi}\right)}^{6}-96G_{6 XXXX} H_{\phi} {H}^{3} {\left(\dot{\pi}\right)}^{7}-720G_{6XXX} H_{\phi} {H}^{3} {\left(\dot{\pi}\right)}^{5}+144G_{6 XXX \pi} H_{\phi} {H}^{2} {\left(\dot{\pi}\right)}^{6}&\nonumber\\
&+48G_{6 XXX \pi} {H}^{3} {\left(\dot{\pi}\right)}^{6}+792G_{6 X X \pi } H_{\phi} {H}^{2} {\left(\dot{\pi}\right)}^{4}+264G_{6 X X \pi } {H}^{3} {\left(\dot{\pi}\right)}^{4}-1080G_{6XX} H_{\phi} {H}^{3} {\left(\dot{\pi}\right)}^{3}&\nonumber\\
&-180G_{6X} H_{\phi} \dot{\pi} {H}^{3}+756G_{6X \pi} H_{\phi} {H}^{2} {\left(\dot{\pi}\right)}^{2}+252G_{6X \pi} {H}^{3} {\left(\dot{\pi}\right)}^{2}+18G_{6\pi} {H}^{3}+54G_{6\pi} H_{\phi} {H}^{2},&\nonumber
\end{flalign}
\begin{flalign}
  &A_{12} = 2F_{X} \dot{\pi}+2K_{X} {\left(\dot{\pi}\right)}^{2}\left(H_{\phi}+3H\right) -2K_{\pi} \dot{\pi}+12G_{4X} H H_{\phi} \dot{\pi}+12G_{4X} \dot{\pi} {H}^{2}&\nonumber\\
  &+2G_{4\pi\pi} \dot{\pi}-4G_{4X \pi} {\left(\dot{\pi}\right)}^{2}\left(5H+2H_{\phi}\right) -2G_{4\pi} H+24G_{4XX} H H_{\phi} {\left(\dot{\pi}\right)}^{3}+24G_{4XX} {H}^{2} {\left(\dot{\pi}\right)}^{3}&\nonumber\\
  &+6G_{5X} {H}^{3} {\left(\dot{\pi}\right)}^{2}+18G_{5X} H_{\phi} {H}^{2} {\left(\dot{\pi}\right)}^{2}-6G_{5\pi} \dot{\pi} {H}^{2}+\left(2H+H_{\phi}\right) G_{5\pi\pi} {\left(\dot{\pi}\right)}^{2}-8G_{5X \pi} {H}^{2} {\left(\dot{\pi}\right)}^{3}&\\
  &-10G_{5X \pi} H H_{\phi} {\left(\dot{\pi}\right)}^{3}-6G_{5\pi} H H_{\phi} \dot{\pi}+12G_{5XX} H_{\phi} {H}^{2} {\left(\dot{\pi}\right)}^{4}+4G_{5XX} {H}^{3} {\left(\dot{\pi}\right)}^{4} +48G_{6XXX} H_{\phi} {H}^{3} {\left(\dot{\pi}\right)}^{5}&\nonumber\\
  &+144G_{6XX} H_{\phi} {H}^{3} {\left(\dot{\pi}\right)}^{3}-96G_{6 X X \pi } H_{\phi} {H}^{2} {\left(\dot{\pi}\right)}^{4}-24G_{6 X X \pi } {H}^{3} {\left(\dot{\pi}\right)}^{4}+36G_{6X} H_{\phi} \dot{\pi} {H}^{3}&\nonumber\\
  &+24G_{6 X \pi \pi} H H_{\phi} {\left(\dot{\pi}\right)}^{3}+12G_{6 X \pi \pi} {H}^{2} {\left(\dot{\pi}\right)}^{3}-168G_{6X \pi} H_{\phi} {H}^{2} {\left(\dot{\pi}\right)}^{2}-48G_{6X \pi} {H}^{3} {\left(\dot{\pi}\right)}^{2}&\nonumber\\
  &+12G_{6\pi\pi} H H_{\phi} \dot{\pi}+6G_{6\pi\pi} \dot{\pi} {H}^{2}-6G_{6\pi} {H}^{3}-12G_{6\pi} H_{\phi} {H}^{2},&\nonumber
\end{flalign}
\begin{flalign}
  &A_{13} = -4G_{4\pi}+8G_{4X} \dot{\pi}\left(H+H_{\phi}\right) +8G_{4X} \ddot{\pi}+8G_{4X \pi} {\left(\dot{\pi}\right)}^{2}+16G_{4XX} \ddot{\pi} {\left(\dot{\pi}\right)}^{2}&\nonumber\\
  &-4G_{5\pi} \dot{\pi}\left(H+H_{\phi}\right) -2G_{5\pi\pi} {\left(\dot{\pi}\right)}^{2}-4G_{5X \pi} \ddot{\pi} {\left(\dot{\pi}\right)}^{2}-4G_{5\pi} \ddot{\pi}+8G_{5X} \dot{\pi} \ddot{\pi}\left(H+H_{\phi}\right) +8G_{5X} H H_{\phi} {\left(\dot{\pi}\right)}^{2}&\nonumber\\
  &+4G_{5X} \dot{H} {\left(\dot{\pi}\right)}^{2}+4G_{5X} {H}^{2} {\left(\dot{\pi}\right)}^{2}+4G_{5X \pi} {\left(\dot{\pi}\right)}^{3}\left(H+H_{\phi}\right) +8G_{5XX} \ddot{\pi} {\left(\dot{\pi}\right)}^{3}\left(H+H_{\phi}\right) +4G_{5X} \dot{H_{\phi}} {\left(\dot{\pi}\right)}^{2}&\nonumber\\
  &+4G_{5X} {H_{\phi}}^{2} {\left(\dot{\pi}\right)}^{2}-48G_{6 X X \pi } \ddot{\pi} {\left(\dot{\pi}\right)}^{3}\left(H_{\phi}+H\right) +48G_{6 X X \pi } H H_{\phi} {\left(\dot{\pi}\right)}^{4}+96G_{6XXX} H H_{\phi} \ddot{\pi} {\left(\dot{\pi}\right)}^{4}&\\
  &+48G_{6XX} H_{\phi} \dot{H} {\left(\dot{\pi}\right)}^{3}+48G_{6XX} H_{\phi} {H}^{2} {\left(\dot{\pi}\right)}^{3}+48G_{6XX} H \dot{H_{\phi}} {\left(\dot{\pi}\right)}^{3}+48G_{6XX} H {H_{\phi}}^{2} {\left(\dot{\pi}\right)}^{3}&\nonumber\\
  &-24G_{6 X \pi \pi} {\left(\dot{\pi}\right)}^{3}\left(H_{\phi}+H\right) +24G_{6X} \left(H H_{\phi} \ddot{\pi}+H \dot{H_{\phi}} \dot{\pi}+H \dot{\pi} {H_{\phi}}^{2}+H_{\phi} \dot{H} \dot{\pi}+H_{\phi} \dot{\pi} {H}^{2}\right)&\nonumber\\
  &-24 G_{6X \pi} \left(3\left(H_{\phi}+H\right) \dot{\pi} \ddot{\pi}+2H H_{\phi} {\left(\dot{\pi}\right)}^{2}+\dot{H_{\phi}} {\left(\dot{\pi}\right)}^{2}+{H_{\phi}}^{2} {\left(\dot{\pi}\right)}^{2}+\dot{H} {\left(\dot{\pi}\right)}^{2}+{H}^{2} {\left(\dot{\pi}\right)}^{2}\right)&\nonumber\\
  &-12G_{6\pi} \left(\dot{H_{\phi}}+{H_{\phi}}^{2}+\dot{H}+{H}^{2}+H H_{\phi}\right),&\nonumber
\end{flalign}
\begin{flalign}
  &A_{14} = F_{X}+2F_{X X}{\left(\dot{\pi}\right)}^{2}-K_{X \pi}{\left(\dot{\pi}\right)}^{2}-K_{\pi}+ 2K_{X}\dot{\pi}\left(H_{\phi}+3H\right) + 2K_{X X}{\left(\dot{\pi}\right)}^{3}\left(H_{\phi}+3H\right) &\nonumber\\
  &-4G_{4 X X \pi }{\left(\dot{\pi}\right)}^{3}\left(H_{\phi}+3H\right)-6G_{4X \pi}\dot{\pi}\left(H_{\phi}+3H\right)+24G_{4XXX} H H_{\phi} {\left(\dot{\pi}\right)}^{4}&\nonumber\\
  &+24G_{4XXX} {H}^{2} {\left(\dot{\pi}\right)}^{4}+6G_{4X}H \left({H}+  H_{\phi}\right)+48G_{4XX} H {\left(\dot{\pi}\right)}^{2} \left({H}+ H_{\phi}\right)&\nonumber\\
  &-3G_{5\pi} H \left(H_{\phi}+ {H}\right)-6G_{5 X X \pi } H H_{\phi} {\left(\dot{\pi}\right)}^{4}-6G_{5 X X \pi } {H}^{2} {\left(\dot{\pi}\right)}^{4}+264G_{6XXX} H_{\phi} {H}^{3} {\left(\dot{\pi}\right)}^{4}&\nonumber\\
  &+ 14 G_{5XX} {\left(\dot{\pi}\right)}^{3} {H}^{2} \left(3H_{\phi} +{H}\right)+4G_{5XXX} {\left(\dot{\pi}\right)}^{5} {H}^{2}\left(3H_{\phi} +{H}\right)+6G_{5X} {H}^{2}\dot{\pi} \left(3H_{\phi} + H\right)&\nonumber\\
  &-96 G_{6 X X \pi } {\left(\dot{\pi}\right)}^{3}{H}^{2} \left(3H_{\phi} +{H}\right)- 24 G_{6 XXX \pi} {\left(\dot{\pi}\right)}^{5} H^2 \left(3H_{\phi} +{H}\right)+48G_{6 XXXX} H_{\phi} {H}^{3} {\left(\dot{\pi}\right)}^{6}&\\
  &-54G_{6X \pi} \dot{\pi} {H}^{2} \left(3H_{\phi} +{H}\right)+18G_{6X} H_{\phi} {H}^{3}-15G_{5X \pi} H {\left(\dot{\pi}\right)}^{2}(H_{\phi}+ {H})&\nonumber\\
  &+252G_{6XX} H_{\phi} {H}^{3}{\left(\dot{\pi}\right)}^{2}, &\nonumber
\end{flalign}

\begin{flalign}
  &A_{15} = -F_{X}+K_{\pi}-K_{X \pi} {\left(\dot{\pi}\right)}^{2}-2K_{X X} \ddot{\pi} {\left(\dot{\pi}\right)}^{2}-2K_{X} \ddot{\pi}-2\left(H_{\phi}+2H\right) K_{X} \dot{\pi}+6G_{4X \pi} \ddot{\pi}&\nonumber\\
  &+6\left(H_{\phi}+2H\right) G_{4X \pi} \dot{\pi}+4G_{4 X X \pi } \ddot{\pi} {\left(\dot{\pi}\right)}^{2}-4\left(H_{\phi}+2H\right) G_{4 X X \pi } {\left(\dot{\pi}\right)}^{3}-8\left(H_{\phi}+2H\right) G_{4XXX} \ddot{\pi} {\left(\dot{\pi}\right)}^{3}&\nonumber\\
  &+2G_{4 X \pi \pi} {\left(\dot{\pi}\right)}^{2}+\left(H_{\phi}+2H\right) G_{5 X \pi \pi} {\left(\dot{\pi}\right)}^{3}-4G_{4XX} \left(3\left(2H+H_{\phi}\right) \dot{\pi} \ddot{\pi}\right.&\nonumber\\
  &\left.+ {\left(\dot{\pi}\right)}^{2} \left(5{H}^{2} +2\dot{H} +\dot{H_{\phi}} +{H_{\phi}}^{2} +6H H_{\phi} \right)\right)-2G_{4X} \left(2\dot{H}+3{H}^{2}+2H H_{\phi}+\dot{H_{\phi}}+{H_{\phi}}^{2}\right)&\nonumber\\
  &+G_{5\pi} \left(2\dot{H}+3{H}^{2}+2H H_{\phi}+\dot{H_{\phi}}+{H_{\phi}}^{2}\right)+G_{5X \pi} \left({\left(\dot{\pi}\right)}^{2}\left(2\dot{H} +5{H}^{2} +6H H_{\phi}+\dot{H_{\phi}} +{H_{\phi}}^{2} \right) \right.&\nonumber\\
  &\left.+\left(4H_{\phi}+8H\right) \dot{\pi} \ddot{\pi}\right)-G_{5XX} \left(10 H {\left(\dot{\pi}\right)}^{2}\ddot{\pi}\left(2H_{\phi} + {H}\right)+ 4 {\left(\dot{\pi}\right)}^{3} \left(12H_{\phi} {H}^{2} +\left(H_{\phi}+H\right) \dot{H}\right.\right.&\nonumber\\
  &\left.\left. +{H}^{3} +H \dot{H_{\phi}} +H {H_{\phi}}^{2} \right)\right)-2G_{5X} \left(\left(2H+2H_{\phi}\right) \dot{H} \dot{\pi}+2\dot{\pi} {H}^{3}+5H_{\phi} \dot{\pi} {H}^{2}+2H \dot{H_{\phi}} \dot{\pi}\right.&\\
  &\left.+2H \dot{\pi} {H_{\phi}}^{2}+2H H_{\phi} \ddot{\pi}+\ddot{\pi} {H}^{2}\right)+2G_{5 X X \pi } \left(\left(H_{\phi}+2H\right) \ddot{\pi} {\left(\dot{\pi}\right)}^{3}-2H H_{\phi} {\left(\dot{\pi}\right)}^{4}-{H}^{2} {\left(\dot{\pi}\right)}^{4}\right)&\nonumber\\
  &-4G_{5XXX} H \ddot{\pi} {\left(\dot{\pi}\right)}^{4}\left(2H_{\phi}+{H}\right)+24G_{6 X X \pi } \left(3 H \ddot{\pi} {\left(\dot{\pi}\right)}^{2}\left(2 H_{\phi} + {H}\right)\right.&\nonumber\\
  &\left. + {\left(\dot{\pi}\right)}^{3}\left(H {H_{\phi}}^{2} +2H_{\phi} {H}^{2} +\left(H_{\phi}+H\right) \dot{H} +{H}^{3} +H \dot{H_{\phi}} \right)\right)+24G_{6 XXX \pi} H {\left(\dot{\pi}\right)}^{4}\left(2 H_{\phi} \ddot{\pi} +\ddot{\pi} {H} - H_{\phi} {H} \dot{\pi}\right)&\nonumber\\
  &-48G_{6 XXXX} H_{\phi} \ddot{\pi} {H}^{2} {\left(\dot{\pi}\right)}^{5}-24G_{6XXX} {H} {\left(\dot{\pi}\right)}^{3}\left(8H_{\phi} \ddot{\pi} H+2 H_{\phi} \dot{H} \dot{\pi}+2H_{\phi} {H}^{2} \dot{\pi}+\dot{H_{\phi}} {H}\dot{\pi}+{H} {H_{\phi}}^{2} \dot{\pi}\right)&\nonumber\\
  &+12 G_{6 XX \pi \pi}H{\left(\dot{\pi}\right)}^{4} \left(2H_{\phi} +{H}\right)-12 G_{6XX} H \left({\left(\dot{\pi}\right)}^{2}\left(8 H_{\phi} \dot{H} +8H_{\phi} {H}^{2} +4{H} {H_{\phi}}^{2}+4\dot{H_{\phi}} {H} \right)+9 H_{\phi} \dot{\pi} \ddot{\pi} {H}\right)&\nonumber\\
  &+18G_{6X \pi} \left(2H H_{\phi} \ddot{\pi}+\ddot{\pi} {H}^{2}+2\left(H+H_{\phi}\right) \dot{H} \dot{\pi}+2\dot{\pi} {H}^{3}+2H \dot{\pi} {H_{\phi}}^{2}+5H_{\phi} \dot{\pi} {H}^{2}+2H \dot{H_{\phi}} \dot{\pi}\right)&\nonumber\\
  &+6 G_{6 X \pi \pi} H {\left(\dot{\pi}\right)}^{2}\left(2 H_{\phi} +{H}\right)-6G_{6X} H \left(\dot{H_{\phi}} {H}+{H} {H_{\phi}}^{2}+2H_{\phi} \dot{H}+2H_{\phi} {H}^{2}\right),&\nonumber
\end{flalign}
\begin{flalign}
  &A_{17} = F_{\pi}-2F_{X \pi} {\left(\dot{\pi}\right)}^{2}+6G_{4\pi} {H}^{2}+2\left(H_{\phi}+3H\right) G_{4\pi\pi} \dot{\pi}+4\left(3H+H_{\phi}\right) G_{4 X \pi \pi} {\left(\dot{\pi}\right)}^{3}&\nonumber\\
  &+6G_{5 X \pi \pi} H H_{\phi} {\left(\dot{\pi}\right)}^{4}+6G_{5 X \pi \pi} {H}^{2} {\left(\dot{\pi}\right)}^{4}+18G_{6\pi\pi} \dot{\pi} {H}^{3}+54G_{6\pi\pi} H_{\phi} \dot{\pi} {H}^{2}+K_{\pi \pi} {\left(\dot{\pi}\right)}^{2}&\nonumber\\
  &-2\left(H_{\phi}+3H\right) K_{X \pi} {\left(\dot{\pi}\right)}^{3}+6G_{4\pi} H H_{\phi}+9G_{5\pi\pi} H H_{\phi} {\left(\dot{\pi}\right)}^{2}+9G_{5\pi\pi} {H}^{2} {\left(\dot{\pi}\right)}^{2}&\nonumber\\
  &-24 G_{4X \pi}{H}{\left(\dot{\pi}\right)}^{2} \left(H+ H_{\phi}\right)-24G_{4 X X \pi } H {\left(\dot{\pi}\right)}^{4}\left(H_{\phi}+{H}\right)-10 G_{5X \pi} {H}^{2} {\left(\dot{\pi}\right)}^{3}\left(3H_{\phi} +{H}\right)&\\
  &-4 G_{5 X X \pi } {\left(\dot{\pi}\right)}^{5} H^2 \left(3 H_{\phi}+{H}\right)-216G_{6 X X \pi } H_{\phi} {H}^{3} {\left(\dot{\pi}\right)}^{4}-108G_{6X \pi} H_{\phi} {H}^{3} {\left(\dot{\pi}\right)}^{2}&\nonumber\\
&+72 G_{6 X \pi \pi} {H}^{2} {\left(\dot{\pi}\right)}^{3}\left(3H_{\phi} +{H}\right)+18G_{6\pi} H_{\phi} {H}^{3}+24 G_{6 XX \pi \pi} {\left(\dot{\pi}\right)}^{5} H^2 \left(3H_{\phi} +{H}\right)&\nonumber\\
&-48G_{6 XXX \pi} H_{\phi} {H}^{3} {\left(\dot{\pi}\right)}^{6},&\nonumber
\end{flalign}
\begin{flalign}
  &A_{18} = -6F_{X} \dot{\pi}-6K_{X \pi} {\left(\dot{\pi}\right)}^{3}+6K_{\pi} \dot{\pi}-12K_{X X} \ddot{\pi} {\left(\dot{\pi}\right)}^{3}-12K_{X} \dot{\pi} \ddot{\pi}-12\left(3H+H_{\phi}\right) K_{X} {\left(\dot{\pi}\right)}^{2}&\nonumber\\
 &+6\left(4H+H_{\phi}\right) G_{4\pi}+12G_{4 X \pi \pi} {\left(\dot{\pi}\right)}^{3}-48\left(H_{\phi}+2H\right) G_{4XXX} \ddot{\pi} {\left(\dot{\pi}\right)}^{4}+6\left(H_{\phi}+2H\right) G_{5 X \pi \pi} {\left(\dot{\pi}\right)}^{4}&\nonumber\\
 &+3\left(H_{\phi}+2H\right) G_{5\pi\pi} {\left(\dot{\pi}\right)}^{2}-12G_{4X} \left(\dot{H_{\phi}} \dot{\pi}+\dot{\pi} {H_{\phi}}^{2}+2\dot{H} \dot{\pi}+9\dot{\pi} {H}^{2}+8H H_{\phi} \dot{\pi}+\left(2H+H_{\phi}\right) \ddot{\pi}\right)&\nonumber\\
 &+12G_{4X \pi} \left(\left(6H+2H_{\phi}\right) {\left(\dot{\pi}\right)}^{2}+3\dot{\pi} \ddot{\pi}\right)-24G_{4XX} \left(9{H}^{2} {\left(\dot{\pi}\right)}^{3}+8H H_{\phi} {\left(\dot{\pi}\right)}^{3}+4\left(H_{\phi}+2H\right) \ddot{\pi} {\left(\dot{\pi}\right)}^{2}\right.&\nonumber\\
 &\left. +2\dot{H} {\left(\dot{\pi}\right)}^{3}+\dot{H_{\phi}} {\left(\dot{\pi}\right)}^{3}+{H_{\phi}}^{2} {\left(\dot{\pi}\right)}^{3}\right)+24G_{4 X X \pi } \left(\ddot{\pi} {\left(\dot{\pi}\right)}^{3}-\left(H_{\phi}+2H\right) {\left(\dot{\pi}\right)}^{4}\right)&\nonumber\\
 &+6G_{5X \pi} \left({\left(\dot{\pi}\right)}^{3} \left(7{H}^{2} +2\dot{H}+\dot{H_{\phi}}+{H_{\phi}}^{2}+4H H_{\phi}\right)+5\left(H_{\phi}+2H\right) \ddot{\pi} {\left(\dot{\pi}\right)}^{2}\right)&\\
 &+6G_{5\pi} \left(\dot{\pi}\left(9 {H}^{2}+2\dot{H} +\dot{H_{\phi}} + {H_{\phi}}^{2}+8H H_{\phi}\right)+\left(H_{\phi}+2H\right) \ddot{\pi}\right)&\nonumber\\
 &-36 G_{5X} \left({\left(\dot{\pi}\right)}^{2}\left(\left(H+H_{\phi}\right) \dot{H} +2{H}^{3} +5H_{\phi} {H}^{2} +H {H_{\phi}}^{2} +H \dot{H_{\phi}} \right) +\dot{\pi} \ddot{\pi} {H}\left(H+2H_{\phi}\right)\right)&\nonumber\\
 &+12G_{5 X X \pi } \left(H_{\phi}+2H\right) \ddot{\pi} {\left(\dot{\pi}\right)}^{4}-12G_{5XX} \left(7\ddot{\pi} {H}^{2} {\left(\dot{\pi}\right)}^{3}+10H_{\phi} {H}^{2} {\left(\dot{\pi}\right)}^{4}+14H H_{\phi} \ddot{\pi} {\left(\dot{\pi}\right)}^{3}\right.&\nonumber\\
 &\left.+2\left(H+H_{\phi}\right) \dot{H} {\left(\dot{\pi}\right)}^{4}+4{H}^{3} {\left(\dot{\pi}\right)}^{4}+2H \dot{H_{\phi}} {\left(\dot{\pi}\right)}^{4}+2H {H_{\phi}}^{2} {\left(\dot{\pi}\right)}^{4}\right)-12G_{5 X X \pi } {\left(\dot{\pi}\right)}^{5} H\left({H}+2 H_{\phi}\right)&\nonumber\\
 &-24G_{5XXX} {\left(\dot{\pi}\right)}^{5} \ddot{\pi} {H}\left({H}+2 H_{\phi}\right)+144G_{6 XXX \pi} {\left(\dot{\pi}\right)}^{5} \ddot{\pi} {H}\left(2 H_{\phi}+ {H}\right)&\nonumber\\
 &+72G_{6 X X \pi } \left(16H H_{\phi} \ddot{\pi} {\left(\dot{\pi}\right)}^{3}+8\ddot{\pi} {H}^{2} {\left(\dot{\pi}\right)}^{3}+5H_{\phi} {H}^{2} {\left(\dot{\pi}\right)}^{4}+2H {H_{\phi}}^{2} {\left(\dot{\pi}\right)}^{4}+2H \dot{H_{\phi}} {\left(\dot{\pi}\right)}^{4}\right.&\nonumber\\
 &\left.+2\left(H_{\phi}+H\right) \dot{H} {\left(\dot{\pi}\right)}^{4}+4{H}^{3} {\left(\dot{\pi}\right)}^{4}\right)-144G_{6 XXX \pi} H_{\phi} {H}^{2} {\left(\dot{\pi}\right)}^{6}-288G_{6 XXXX} H_{\phi} \ddot{\pi} {H}^{2} {\left(\dot{\pi}\right)}^{6}&\nonumber\\
 &-1584G_{6XXX} H_{\phi} \ddot{\pi} {H}^{2} {\left(\dot{\pi}\right)}^{4}-144G_{6XXX} {\left(\dot{\pi}\right)}^{5} H \left(2H_{\phi} \dot{H}+4H_{\phi} {H}^{2}+\dot{H_{\phi}} {H}+{H} {H_{\phi}}^{2}\right)&\nonumber\\
 &-216G_{6XX} H  {\left(\dot{\pi}\right)}^{2}\left(8H_{\phi} {H}^{3} \dot{\pi}+7H_{\phi} \ddot{\pi} {H}^{2} +2{H}^{2} {H_{\phi}}^{2} \dot{\pi}+2\dot{H_{\phi}} {H}^{2} \dot{\pi}+4H H_{\phi} \dot{H} \dot{\pi}\right)&\nonumber\\
 &-108G_{6X} \left(H_{\phi} \ddot{\pi} {H}^{2}+\dot{H_{\phi}} \dot{\pi} {H}^{2}+\dot{\pi} {H}^{2} {H_{\phi}}^{2}+2H H_{\phi} \dot{H} \dot{\pi}+4H_{\phi} \dot{\pi} {H}^{3}\right)&\nonumber\\
 &+36 G_{6X \pi} \dot{\pi}\left(18 H H_{\phi}\ddot{\pi}+9\ddot{\pi} {H}^{2}+8H {H_{\phi}}^{2} \dot{\pi}+39H_{\phi} {H}^{2} \dot{\pi}+8H \dot{H_{\phi}} \dot{\pi}+8\left(H_{\phi}+H\right) \dot{H} \dot{\pi}+16{H}^{3} \dot{\pi}\right)&\nonumber\\
 &+72G_{6 XX \pi \pi} {\left(\dot{\pi}\right)}^{5}H \left(2H_{\phi}+{H}\right)+108 G_{6 X \pi \pi}H {\left(\dot{\pi}\right)}^{3}\left(2 H_{\phi} +{H}\right)&\nonumber\\
 &+18G_{6\pi} \left(2H \dot{H_{\phi}}+2H {H_{\phi}}^{2}+2\left(H+H_{\phi}\right) \dot{H}+2{H}^{3}+9H_{\phi} {H}^{2}\right),&\nonumber
\end{flalign}
\begin{flalign}
 & A_{20} = \frac{1}{2}F_{\pi \pi}-K_{X \pi} \left({\left(\dot{\pi}\right)}^{2} \left(9{H}^{2}+3\dot{H}+\dot{H_{\phi}}+{H_{\phi}}^{2}+6H H_{\phi}\right)+2\dot{\pi} \ddot{\pi} \left(H_{\phi}+3H\right)\right)&\nonumber\\
 &-6 G_{4X \pi} \left(\dot{\pi} \left(\dot{H} \left(2H+H_{\phi}\right)+3{H}^{3}+H \dot{H_{\phi}}+H {H_{\phi}}^{2}+4H_{\phi} {H}^{2}\right)+\ddot{\pi} H\left({H}+ H_{\phi}\right)\right)&\nonumber\\
 &+2G_{4 X \pi \pi} \left(3\dot{\pi} \ddot{\pi} \left(H_{\phi}+3H\right)+{\left(\dot{\pi}\right)}^{2} \left(6H H_{\phi}+9{H}^{2}+6\dot{H}+\dot{H_{\phi}}+{H_{\phi}}^{2}\right)\right)&\nonumber\\
 &+4G_{4 XX \pi \pi} \left(\ddot{\pi} {\left(\dot{\pi}\right)}^{3} \left(H_{\phi}+3H\right)- 3{\left(\dot{\pi}\right)}^{4} H\left( H_{\phi}+{H}\right)\right)-24 G_{4 XXX \pi} H\ddot{\pi} {\left(\dot{\pi}\right)}^{4} \left(H_{\phi}+{H}\right)&\nonumber\\
 &-12 G_{4 X X \pi } \left(4\ddot{\pi} {\left(\dot{\pi}\right)}^{2} H \left({H}+ H_{\phi}\right)+{\left(\dot{\pi}\right)}^{3} \left(4H_{\phi} {H}^{2}+\dot{H} \left(H_{\phi}+2H\right)+3{H}^{3}+H \dot{H_{\phi}}+H {H_{\phi}}^{2}\right)\right)&\nonumber\\
  &+G_{4\pi\pi} \left(\dot{H_{\phi}}+{H_{\phi}}^{2}+3\dot{H}+6{H}^{2}+3H H_{\phi}\right)+3 G_{5\pi\pi} \left(\dot{\pi} \left(H {H_{\phi}}^{2}+\dot{H} \left(2H+H_{\phi}\right)+3{H}^{3}+H \dot{H_{\phi}}\right.\right.&\nonumber\\
  &\left.\left.+4H_{\phi} {H}^{2}\right)+\ddot{\pi} H \left( H_{\phi}+{H}\right)\right)+G_{5 X \pi \pi} \left({\left(\dot{\pi}\right)}^{3} \left(3H {H_{\phi}}^{2}+\dot{H} \left(6H+3H_{\phi}\right)+7{H}^{3}+6H_{\phi} {H}^{2}+3H \dot{H_{\phi}}\right)\right.&\nonumber\\
 &\left.+15\ddot{\pi} {\left(\dot{\pi}\right)}^{2}H \left( H_{\phi}+{H}\right)\right)-G_{5 X X \pi } \left(14\ddot{\pi} {\left(\dot{\pi}\right)}^{3} H^2 \left(3H_{\phi} +{H}\right)+2{\left(\dot{\pi}\right)}^{4} H \left(10H_{\phi} {H}^{2}+6 H_{\phi} \dot{H}+3\dot{H} {H} \right.\right.&\nonumber\\
 &\left.\left.+3{H}^{3}+3\dot{H_{\phi}} {H}+3{H} {H_{\phi}}^{2}\right)\right)-3G_{5X \pi} \left({\left(\dot{\pi}\right)}^{2} H \left(3{H} {H_{\phi}}^{2}+3\dot{H} {H}+3{H}^{3}+10 H_{\phi} {H}^{2}+3\dot{H_{\phi}} {H}\right.\right.&\nonumber\\
 &\left.\left. +6 H_{\phi} \dot{H}\right) +2\dot{\pi} \ddot{\pi} H^2 \left(3H_{\phi} +{H}\right)\right)+\frac{3}{2}G_{5 \pi \pi \pi} {\left(\dot{\pi}\right)}^{2} H \left( H_{\phi}+{H}\right)+2 G_{5 XX \pi \pi} \left(3\ddot{\pi} {\left(\dot{\pi}\right)}^{4}H \left(H_{\phi}+{H}\right)\right.&\nonumber\\
 &\left. -{\left(\dot{\pi}\right)}^{5} H^2 \left(3H_{\phi}+{H}\right)\right)-4G_{5 XXX \pi} \ddot{\pi} {\left(\dot{\pi}\right)}^{5} H^2 \left(3H_{\phi} +{H}\right)+3G_{5 X \pi \pi \pi} {\left(\dot{\pi}\right)}^{4} H \left(H_{\phi}+{H}\right)&\nonumber\\
 &+G_{6 XX \pi \pi} \left(96 \ddot{\pi} {\left(\dot{\pi}\right)}^{3} H^2 \left(3H_{\phi}+{H}\right) +12{\left(\dot{\pi}\right)}^{4} H \left(3{H} {H_{\phi}}^{2}+5H_{\phi} {H}^{2}+6 H_{\phi} \dot{H}\right.\right.&\nonumber\\
 &\left.\left.+3\dot{H} {H}+3{H}^{3}+3\dot{H_{\phi}} {H}\right)\right) +24 G_{6 XXX \pi \pi} \left(\ddot{\pi} {\left(\dot{\pi}\right)}^{5} H^2 \left(3H_{\phi} +{H}\right)-H_{\phi} {H}^{3} {\left(\dot{\pi}\right)}^{6}\right)&\nonumber\\
 &-24G_{6 XXX \pi} \left(11H_{\phi} \ddot{\pi} {H}^{3} {\left(\dot{\pi}\right)}^{4} +{\left(\dot{\pi}\right)}^{5} H^2 \left(3H_{\phi} \dot{H} +3H_{\phi} {H}^{2}+\dot{H_{\phi}} {H}+{H} {H_{\phi}}^{2}\right)\right)&\nonumber\\
 &+12 G_{6 XX \pi \pi \pi} {\left(\dot{\pi}\right)}^{5} H^2\left(3H_{\phi}+{H}\right)-36 G_{6 X X \pi } \left(7H_{\phi} \ddot{\pi} {H}^{3} {\left(\dot{\pi}\right)}^{2}\right.&\\
 &\left.+{\left(\dot{\pi}\right)}^{3} \left(2{H}^{3} {H_{\phi}}^{2}+6H_{\phi} \dot{H} {H}^{2}+6H_{\phi} {H}^{4}+2\dot{H_{\phi}} {H}^{3}\right)\right)+G_{6 X \pi \pi} \left(54 \dot{\pi} \ddot{\pi}H^2 \left(3H_{\phi} +{H}\right)\right.&\nonumber\\
 &\left. +18 {\left(\dot{\pi}\right)}^{2} \left(4\dot{H_{\phi}} {H}^{2}+4{H}^{2} {H_{\phi}}^{2}+144H H_{\phi} \dot{H}+13H_{\phi} {H}^{3}+4\dot{H} {H}^{2}+4{H}^{4}\right)\right)&\nonumber\\
 &+18 G_{6 X \pi \pi \pi} {\left(\dot{\pi}\right)}^{3} H^2 \left(3H_{\phi}+{H}\right)-18 G_{6X \pi} H^2 \left(\dot{\pi} \left(3H_{\phi} \dot{H}+3H_{\phi} {H}^{2}+{H} {H_{\phi}}^{2}+\dot{H_{\phi}} {H}\right)\right.&\nonumber\\
 &\left. +H_{\phi} H \ddot{\pi} \right) +9 G_{6\pi\pi} H \left(\dot{H_{\phi}} {H}+{H} {H_{\phi}}^{2}+\dot{H} {H}+2 H_{\phi} \dot{H}+{H}^{3}+3H_{\phi} {H}^{2}\right)&\nonumber\\
 &-F_{X \pi \pi}{\left(\dot{\pi}\right)}^{2}+\frac{1}{2}K_{\pi \pi \pi}{\left(\dot{\pi}\right)}^{2}-2F_{X X \pi}\ddot{\pi} {\left(\dot{\pi}\right)}^{2} -F_{X \pi} \left(\ddot{\pi}+  \dot{\pi}\left(H_{\phi}+3H\right) \right)&\nonumber\\
 & +K_{\pi \pi}\left(\ddot{\pi}+ \dot{\pi}\left(H_{\phi}+3H\right)\right) +K_{X \pi \pi}\left(\ddot{\pi} {\left(\dot{\pi}\right)}^{2}- {\left(\dot{\pi}\right)}^{3} \left(H_{\phi}+3H\right)\right)+2 G_{4 X \pi \pi \pi} {\left(\dot{\pi}\right)}^{3} \left(H_{\phi}+3H\right)&\nonumber\\
 &-2K_{XX \pi} \ddot{\pi} {\left(\dot{\pi}\right)}^{3} \left(H_{\phi}+3H\right)-48G_{6 XXXX \pi} H_{\phi} \ddot{\pi} {H}^{3} {\left(\dot{\pi}\right)}^{6},&\nonumber
 \end{flalign}
 \begin{flalign}
   &B_1 = \frac{1}{2}{G_{4}}-G_{4X} {\left(\dot{\pi}\right)}^{2}+\frac{1}{4}G_{5\pi} {\left(\dot{\pi}\right)}^{2}-G_{5X} H {\left(\dot{\pi}\right)}^{3}-6G_{6XX} {H}^{2} {\left(\dot{\pi}\right)}^{4}+6G_{6X \pi} H {\left(\dot{\pi}\right)}^{3}&\nonumber\\
   &-6G_{6X} {H}^{2} {\left(\dot{\pi}\right)}^{2}+\frac{3}{2}{G_{6}} {H}^{2}+3G_{6\pi} H \dot{\pi},&
 \end{flalign}
 \begin{flalign}
   &B_2 = -{G_{4}} \left(3H_{\phi}+H\right)+2G_{4X} \left({\left(\dot{\pi}\right)}^{2} \left(3H_{\phi}+H\right)+\dot{\pi} \ddot{\pi}\right)+G_{5X \pi} {\left(\dot{\pi}\right)}^{3}\left(2H \dot{\pi} -\ddot{\pi}\right)&\nonumber\\
   &- \dfrac{1}{2} G_{5\pi} \left(2\dot{\pi} \ddot{\pi}+{\left(\dot{\pi}\right)}^{2} \left( H +3H_{\phi}\right)\right)+2G_{5X} \left(3H \ddot{\pi} {\left(\dot{\pi}\right)}^{2}+{\left(\dot{\pi}\right)}^{3} \left(3H H_{\phi}+\dot{H}+{H}^{2}\right)\right)&\nonumber\\
   &+12G_{6XX} {\left(\dot{\pi}\right)}^{3} H\left(6\ddot{\pi} {H} +\dot{\pi} \left(2\dot{H}+{H}^{2}+3H_{\phi} H\right)\right)+12 G_{6 X X \pi } H{\left(\dot{\pi}\right)}^{4} \left({H} \dot{\pi}-2\ddot{\pi} \right)&\\
   &+6G_{6X} \dot{\pi} H \left(3 \ddot{\pi} {H}+2\dot{\pi} \left(2\dot{H}+{H}^{2}+3H_{\phi} H\right)\right)-12 G_{6X \pi} {\left(\dot{\pi}\right)}^{2}\left(4H \ddot{\pi} \right.&\nonumber\\
   &\left.+\dot{\pi} \left(3H H_{\phi}+\dot{H}\right)\right)-3{G_{6}} H\left(2\dot{H}+{H}^{2}+3H_{\phi} H\right)-3 G_{6\pi} \left(\dot{\pi} \left(3{H}^{2}+2\dot{H}+6H H_{\phi}\right)\right.&\nonumber\\
   &\left. +2H \ddot{\pi}\right)-G_{4\pi} \dot{\pi}+ 2G_{4X \pi}{\left(\dot{\pi}\right)}^{3} - \frac{1}{2}G_{5\pi\pi}{\left(\dot{\pi}\right)}^{3}+4G_{4XX} \ddot{\pi} {\left(\dot{\pi}\right)}^{3}+4G_{5XX} H \ddot{\pi} {\left(\dot{\pi}\right)}^{4}&\nonumber\\
   &+24G_{6XXX} \ddot{\pi} {H}^{2} {\left(\dot{\pi}\right)}^{5}-12G_{6 X \pi \pi} H {\left(\dot{\pi}\right)}^{4}-6G_{6\pi\pi} H {\left(\dot{\pi}\right)}^{2},&\nonumber
 \end{flalign}
 \begin{flalign}
   &C_1 = -12 B_1,&
 \end{flalign}
\begin{flalign}
 &C_2 = (H_{\phi} - H) C_1,&\nonumber
\end{flalign}
\begin{flalign}
  &C_4 = -2K_{X} {\left(\dot{\pi}\right)}^{2}+4G_{4X \pi} {\left(\dot{\pi}\right)}^{2}+2G_{4\pi}-24G_{4XX} H {\left(\dot{\pi}\right)}^{3}-12G_{4X} H \dot{\pi}+6G_{5X \pi} H {\left(\dot{\pi}\right)}^{3}&\nonumber\\
  &+6G_{5\pi} H \dot{\pi}-18G_{5X} {H}^{2} {\left(\dot{\pi}\right)}^{2}-12G_{5XX} {H}^{2} {\left(\dot{\pi}\right)}^{4}-144G_{6XX} {H}^{3} {\left(\dot{\pi}\right)}^{3}+72G_{6 X X \pi } {H}^{2} {\left(\dot{\pi}\right)}^{4}&\\
  &-48G_{6XXX} {H}^{3} {\left(\dot{\pi}\right)}^{5}-36G_{6X} \dot{\pi} {H}^{3}+144G_{6X \pi} {H}^{2} {\left(\dot{\pi}\right)}^{2}+18G_{6\pi} {H}^{2},&\nonumber
\end{flalign}
\begin{flalign}
  &C_5 = -2F_{X}\dot{\pi}+2K_{\pi}\dot{\pi}-4K_{X X} \ddot{\pi} {\left(\dot{\pi}\right)}^{3}-2K_{X \pi}{\left(\dot{\pi}\right)}^{3}+4G_{4 X \pi \pi}{\left(\dot{\pi}\right)}^{3}-48G_{4XXX} H \ddot{\pi} {\left(\dot{\pi}\right)}^{4}&\nonumber\\
  &+2G_{4\pi} \left(2H_{\phi}+3H\right)-4K_{X} \left({\left(\dot{\pi}\right)}^{2} \left(H_{\phi}+3H\right)+\dot{\pi} \ddot{\pi}\right)-12 G_{4X} \left(\dot{\pi} \left(4{H}^{2}+2H H_{\phi}+\dot{H}\right)\right.&\nonumber\\
  &\left. +H \ddot{\pi}\right)+4G_{4X \pi} \left(2 {\left(\dot{\pi}\right)}^{2} \left(H_{\phi}+3H\right)+3\dot{\pi} \ddot{\pi}\right)+8G_{4 X X \pi } {\left(\dot{\pi}\right)}^{3} \left(\ddot{\pi}-3H \dot{\pi}\right)&\nonumber\\
  &-24 G_{4XX} \left({\left(\dot{\pi}\right)}^{3} \left(2H H_{\phi}+4{H}^{2}+\dot{H}\right)+4H \ddot{\pi} {\left(\dot{\pi}\right)}^{2}\right)+6 G_{5X \pi} {\left(\dot{\pi}\right)}^{2} \left(\dot{\pi} \left(2{H}^{2}+2H H_{\phi}+\dot{H}\right)\right.&\nonumber\\
  &\left.+5H \ddot{\pi} \right)+6G_{5\pi} \left(\dot{\pi} \left(4{H}^{2}+2H H_{\phi}+\dot{H}\right)+H \ddot{\pi}\right)-4 G_{5XX} {H} {\left(\dot{\pi}\right)}^{3}\left(2\dot{\pi} \left(5{H}^{2}+3H_{\phi} H+3\dot{H}\right)\right.&\nonumber\\
  &\left. +21\ddot{\pi} H\right)-12 G_{5X} \dot{\pi} {H}\left(\dot{\pi} \left(5{H}^{2}+3H_{\phi} H+3\dot{H}\right)+3 \ddot{\pi} {H}\right)+12 G_{5 X X \pi } H {\left(\dot{\pi}\right)}^{4}\left(\ddot{\pi} -{H} \dot{\pi}\right)&\\
  &+24 G_{6 X X \pi } {H} {\left(\dot{\pi}\right)}^{3}\left(\dot{\pi} \left(5{H}^{2}+6H_{\phi} {H}+6 \dot{H}\right)+24\ddot{\pi} H\right)-72 G_{6XX}{H}^{2} {\left(\dot{\pi}\right)}^{2} \left(2 \dot{\pi} \left(2H_{\phi} H\right.\right.&\nonumber\\
  &\left.\left. +3\dot{H}+3{H}^{2}\right)+7\ddot{\pi} H \right)-48 G_{6XXX} {H}^{2} {\left(\dot{\pi}\right)}^{4} \left(11 \ddot{\pi} H +\dot{\pi} \left(2H_{\phi} H+3\dot{H}+3{H}^{2}\right)\right)&\nonumber\\
  &+48 G_{6 XXX \pi} {H}^{2} {\left(\dot{\pi}\right)}^{5} \left(3\ddot{\pi} -{H} \dot{\pi}\right)+36 G_{6X \pi} \dot{\pi}  H \left(\dot{\pi} \left(13 {H}^{2}+8H_{\phi} {H}+8 \dot{H}\right)+9 \ddot{\pi} {H}\right)&\nonumber\\
  & -36 G_{6X} H^2 \left(\dot{\pi}  \left(2H_{\phi} H+3\dot{H}+3{H}^{2}\right)+\ddot{\pi} {H}\right)+18G_{6\pi} H \left(2H_{\phi} H+2\dot{H}+3{H}^{2}\right)&\nonumber\\
  & +3G_{5\pi\pi} H {\left(\dot{\pi}\right)}^{2}+6G_{5 X \pi \pi} H {\left(\dot{\pi}\right)}^{4}-24G_{5XXX} \ddot{\pi} {H}^{2} {\left(\dot{\pi}\right)}^{5}+72G_{6 XX \pi \pi} {H}^{2} {\left(\dot{\pi}\right)}^{5}&\nonumber\\
  &-96G_{6 XXXX} \ddot{\pi} {H}^{3} {\left(\dot{\pi}\right)}^{6}+108G_{6 X \pi \pi} {\left(\dot{\pi}\right)}^{3}{H}^{2}& \nonumber
\end{flalign}
 \begin{flalign}
   &C_7 = - \dfrac13 C_1,&\\
   &C_8 = \dfrac13 C_1 (H - H_{\phi}),&
 \end{flalign}
 \begin{flalign}
   &C_9 = -2K_{X} {\left(\dot{\pi}\right)}^{3}+6{G_{4}} H-24G_{4X} H {\left(\dot{\pi}\right)}^{2}+4G_{4X \pi} {\left(\dot{\pi}\right)}^{3}+2G_{4\pi} \dot{\pi}-24G_{4XX} H {\left(\dot{\pi}\right)}^{4}&\nonumber\\
   &-30G_{5X} {H}^{2} {\left(\dot{\pi}\right)}^{3}+9G_{5\pi} H {\left(\dot{\pi}\right)}^{2}+6G_{5X \pi} H {\left(\dot{\pi}\right)}^{4}-12G_{5XX} {H}^{2} {\left(\dot{\pi}\right)}^{5}-216G_{6XX} {H}^{3} {\left(\dot{\pi}\right)}^{4}&\\
   &-48G_{6XXX} {H}^{3} {\left(\dot{\pi}\right)}^{6}+72G_{6 X X \pi } {H}^{2} {\left(\dot{\pi}\right)}^{5}-108G_{6X} {H}^{3} {\left(\dot{\pi}\right)}^{2}+216G_{6X \pi} {H}^{2} {\left(\dot{\pi}\right)}^{3}&\nonumber\\
   &+18{G_{6}} {H}^{3}+54G_{6\pi} \dot{\pi} {H}^{2}&\nonumber
 \end{flalign}
 \begin{flalign}
   &C_{10} = -2G_{4\pi}+4 G_{4X} \left(2H \dot{\pi}+\ddot{\pi}\right)+ 4G_{4X \pi}{\left(\dot{\pi}\right)}^{2}+8G_{4XX} \ddot{\pi} {\left(\dot{\pi}\right)}^{2}-G_{5\pi\pi}{\left(\dot{\pi}\right)}^{2}&\nonumber\\
   &+8G_{5XX} H \ddot{\pi} {\left(\dot{\pi}\right)}^{3}-2G_{5\pi} \left(2H \dot{\pi}+\ddot{\pi}\right)+G_{5X} \left({\left(\dot{\pi}\right)}^{2} \left(8{H}^{2}+4\dot{H}\right)+8H \dot{\pi} \ddot{\pi}\right)&\nonumber\\
   &+2G_{5X \pi} {\left(\dot{\pi}\right)}^{2} \left( 2H \dot{\pi}-\ddot{\pi} \right)+48 G_{6XX} {H} {\left(\dot{\pi}\right)}^{2}\left(2\ddot{\pi} H +\dot{\pi} \left(\dot{H}+{H}^{2}\right)\right)-24G_{6 X \pi \pi} H {\left(\dot{\pi}\right)}^{3}&\nonumber\\
   &+24 G_{6 X X \pi }{H} {\left(\dot{\pi}\right)}^{3} \left(H \dot{\pi}-2  \ddot{\pi}\right)+12G_{6X} H \left(\ddot{\pi} {H}+2 \dot{\pi} \left( \dot{H}+{H}^{2}\right)\right)&\\
   &-24G_{6X \pi}\dot{\pi} \left(\dot{\pi} \left(\dot{H}+2{H}^{2}\right)+3H \ddot{\pi}\right)-6G_{6\pi} \left(2\dot{H}+3{H}^{2}\right)+48G_{6XXX} \ddot{\pi} {H}^{2} {\left(\dot{\pi}\right)}^{4}&\nonumber
 \end{flalign}
 \begin{flalign}
   &C_{11} = -4{G_{4}}+ 24 G_{6X} \left(H \dot{\pi} \ddot{\pi}+{\left(\dot{\pi}\right)}^{2} \left({H}^{2}+\dot{H}\right)\right)+ 24 G_{6X \pi} {\left(\dot{\pi}\right)}^{2} \left(-\ddot{\pi} +H {\left(\dot{\pi}\right)}\right)&\nonumber\\
   &-12{G_{6}} \left(\dot{H}+{H}^{2}\right) - 12 G_{6\pi} \left(H \dot{\pi}+ \ddot{\pi}\right)+{\left(\dot{\pi}\right)}^{2} \left(2G_{5\pi}-12G_{6\pi\pi}\right)& \nonumber \\
   &+4G_{5X} \ddot{\pi} {\left(\dot{\pi}\right)}^{2}+48G_{6XX} H \ddot{\pi} {\left(\dot{\pi}\right)}^{3},&
\end{flalign}
\begin{flalign}
    &C_{12} = \dfrac13 C_1.&
 \end{flalign}

\bibliographystyle{IEEEtran}
%amsplaini
\bibliography{main}

\begin{thebibliography}{10}
\providecommand{\url}[1]{#1}
\csname url@samestyle\endcsname
\providecommand{\newblock}{\relax}
\providecommand{\bibinfo}[2]{#2}
\providecommand{\BIBentrySTDinterwordspacing}{\spaceskip=0pt\relax}
\providecommand{\BIBentryALTinterwordstretchfactor}{4}
\providecommand{\BIBentryALTinterwordspacing}{\spaceskip=\fontdimen2\font plus
\BIBentryALTinterwordstretchfactor\fontdimen3\font minus
  \fontdimen4\font\relax}
\providecommand{\BIBforeignlanguage}[2]{{%
\expandafter\ifx\csname l@#1\endcsname\relax
\typeout{** WARNING: IEEEtran.bst: No hyphenation pattern has been}%
\typeout{** loaded for the language `#1'. Using the pattern for}%
\typeout{** the default language instead.}%
\else
\language=\csname l@#1\endcsname
\fi
#2}}
\providecommand{\BIBdecl}{\relax}
\BIBdecl

\bibitem{LIGOScientific:2017vwq}
B.~P. Abbott \emph{et~al.}, ``{GW170817: Observation of Gravitational Waves
  from a Binary Neutron Star Inspiral},'' \emph{Phys. Rev. Lett.}, vol. 119,
  no.~16, p. 161101, 2017.

\bibitem{horndeski1974second}
G.~W. Horndeski, ``Second-order scalar-tensor field equations in a
  four-dimensional space,'' \emph{International Journal of Theoretical
  Physics}, vol.~10, no.~6, pp. 363--384, 1974.

\bibitem{nicolis2009galileon}
A.~Nicolis, R.~Rattazzi, and E.~Trincherini, ``Galileon as a local modification
  of gravity,'' \emph{Physical Review D}, vol.~79, no.~6, p. 064036, 2009.

\bibitem{Deffayet:2011gz}
C.~Deffayet, X.~Gao, D.~A. Steer, and G.~Zahariade, ``{From k-essence to
  generalised Galileons},'' \emph{Phys. Rev. D}, vol.~84, p. 064039, 2011.

\bibitem{Abdalla:2022yfr}
E.~Abdalla \emph{et~al.}, ``{Cosmology intertwined: A review of the particle
  physics, astrophysics, and cosmology associated with the cosmological
  tensions and anomalies},'' \emph{JHEAp}, vol.~34, pp. 49--211, 2022.

\bibitem{Bettoni:2016mij}
D.~Bettoni, J.~M. Ezquiaga, K.~Hinterbichler, and M.~Zumalac\'arregui, ``{Speed
  of Gravitational Waves and the Fate of Scalar-Tensor Gravity},'' \emph{Phys.
  Rev. D}, vol.~95, no.~8, p. 084029, 2017.

\bibitem{Ezquiaga:2017ekz}
J.~M. Ezquiaga and M.~Zumalac\'arregui, ``{Dark Energy After GW170817: Dead
  Ends and the Road Ahead},'' \emph{Phys. Rev. Lett.}, vol. 119, no.~25, p.
  251304, 2017.

\bibitem{Sakstein:2017xjx}
J.~Sakstein and B.~Jain, ``{Implications of the Neutron Star Merger GW170817
  for Cosmological Scalar-Tensor Theories},'' \emph{Phys. Rev. Lett.}, vol.
  119, no.~25, p. 251303, 2017.

\bibitem{Baker:2017hug}
T.~Baker, E.~Bellini, P.~G. Ferreira, M.~Lagos, J.~Noller, and I.~Sawicki,
  ``{Strong constraints on cosmological gravity from GW170817 and GRB
  170817A},'' \emph{Phys. Rev. Lett.}, vol. 119, no.~25, p. 251301, 2017.

\bibitem{Creminelli:2017sry}
P.~Creminelli and F.~Vernizzi, ``{Dark Energy after GW170817 and GRB170817A},''
  \emph{Phys. Rev. Lett.}, vol. 119, no.~25, p. 251302, 2017.

\bibitem{Mironov:2024idn}
S.~Mironov, A.~Shtennikova, and M.~Valencia-Villegas, ``{Reviving Horndeski
  after GW170817 by Kaluza-Klein compactifications},'' \emph{Phys. Lett. B},
  vol. 858, p. 139058, 2024.

\bibitem{Pirtskhalava:2015nla}
D.~Pirtskhalava, L.~Santoni, E.~Trincherini, and F.~Vernizzi, ``{Weakly Broken
  Galileon Symmetry},'' \emph{JCAP}, vol.~09, p. 007, 2015.

\bibitem{Goon:2016ihr}
G.~Goon, K.~Hinterbichler, A.~Joyce, and M.~Trodden, ``{Aspects of Galileon
  Non-Renormalization},'' \emph{JHEP}, vol.~11, p. 100, 2016.

\bibitem{MironovNew}
S.~Mironov, M.~Sharov, and V.~Volkova, ``{Time-dependent, spherically symmetric
  background in Kaluza-Klein compactified Horndeski theory and the speed of
  gravity waves},'' 8 2024.

\bibitem{Mironov:2024yqa}
------, ``{Linear stability of a time-dependent, spherically symmetric
  background in beyond Horndeski theory and the speed of gravity waves},'' 8
  2024.

\bibitem{Charmousis:2011bf}
C.~Charmousis, E.~J. Copeland, A.~Padilla, and P.~M. Saffin, ``{General second
  order scalar-tensor theory, self tuning, and the Fab Four},'' \emph{Phys.
  Rev. Lett.}, vol. 108, p. 051101, 2012.

\bibitem{Babichev:2024kfo}
E.~Babichev, C.~Charmousis, B.~Muntz, A.~Padilla, and I.~D. Saltas,
  ``{Horndeski speed tests with scalar-photon couplings},'' 7 2024.

\bibitem{Horndeski:1976gi}
G.~W. Horndeski, ``{Conservation of Charge and the Einstein-Maxwell Field
  Equations},'' \emph{J. Math. Phys.}, vol.~17, pp. 1980--1987, 1976.

\bibitem{Buchdahl:1979wi}
H.~A. Buchdahl, ``{ON A LAGRANGIAN FOR NONMINIMALLY COUPLED GRAVITATIONAL AND
  ELECTROMAGNETIC FIELDS},'' \emph{J. Phys. A}, vol.~12, pp. 1037--1043, 1979.

\bibitem{Mueller-Hoissen:1987nvb}
F.~Mueller-Hoissen, ``{Modification of Einstein {Yang-Mills} Theory From
  Dimensional Reduction of the {Gauss-Bonnet} Action},'' \emph{Class. Quant.
  Grav.}, vol.~5, p. L35, 1988.

\bibitem{Charmousis:2008kc}
C.~Charmousis, ``{Higher order gravity theories and their black hole
  solutions},'' \emph{Lect. Notes Phys.}, vol. 769, pp. 299--346, 2009.

\bibitem{Nejati:2024tuo}
M.~S. Nejati and M.~H. Vahidinia, ``{Two-dimensional (bi-)scalar gravities from
  four-dimensional Horndeski},'' \emph{Class. Quant. Grav.}, vol.~41, no.~22,
  p. 225013, 2024.

\bibitem{Tasinato:2013oja}
G.~Tasinato, K.~Koyama, and N.~Khosravi, ``{The role of vector fields in
  modified gravity scenarios},'' \emph{JCAP}, vol.~11, p. 037, 2013.

\bibitem{Petrov:2018xtx}
P.~Petrov, ``{Galileon-like vector fields},'' \emph{Phys. Rev. D}, vol. 100,
  no.~2, p. 025006, 2019.

\bibitem{Deffayet:2013tca}
C.~Deffayet, A.~E. G\"umr\"uk\c{c}\"uo\u{g}lu, S.~Mukohyama, and Y.~Wang, ``{A
  no-go theorem for generalized vector Galileons on flat spacetime},''
  \emph{JHEP}, vol.~04, p. 082, 2014.

\bibitem{Colleaux:2023cqu}
A.~Coll\'eaux, D.~Langlois, and K.~Noui, ``{Classification of generalised
  higher-order Einstein-Maxwell Lagrangians},'' \emph{JHEP}, vol.~03, p. 041,
  2024.

\bibitem{Colleaux:2024ndy}
------, ``{Degenerate Higher-Order Maxwell Theories in Flat Space-Time},'' 4
  2024.

\bibitem{VanAcoleyen:2011mj}
K.~Van~Acoleyen and J.~Van~Doorsselaere, ``{Galileons from Lovelock actions},''
  \emph{Phys. Rev. D}, vol.~83, p. 084025, 2011.

\bibitem{deRham:2010eu}
C.~de~Rham and A.~J. Tolley, ``{DBI and the Galileon reunited},'' \emph{JCAP},
  vol.~05, p. 015, 2010.

\bibitem{Trodden:2011xh}
M.~Trodden and K.~Hinterbichler, ``{Generalizing Galileons},'' \emph{Class.
  Quant. Grav.}, vol.~28, p. 204003, 2011.

\bibitem{Easson:2020bgk}
D.~Easson, T.~Manton, M.~Parikh, and A.~Svesko, ``{The Stringy Origins of
  Galileons and their Novel Limit},'' \emph{JCAP}, vol.~05, p. 031, 2021.

\bibitem{Charmousis:2014mia}
C.~Charmousis, ``{From Lovelock to Horndeski`s Generalized Scalar Tensor
  Theory},'' \emph{Lect. Notes Phys.}, vol. 892, pp. 25--56, 2015.

\bibitem{Deffayet:2009mn}
C.~Deffayet, S.~Deser, and G.~Esposito-Farese, ``{Generalized Galileons: All
  scalar models whose curved background extensions maintain second-order field
  equations and stress-tensors},'' \emph{Phys. Rev. D}, vol.~80, p. 064015,
  2009.

\bibitem{Esposito-Farese:2009wbc}
G.~Esposito-Farese, C.~Pitrou, and J.-P. Uzan, ``{Vector theories in
  cosmology},'' \emph{Phys. Rev. D}, vol.~81, p. 063519, 2010.

\bibitem{Golovnev:2008cf}
A.~Golovnev, V.~Mukhanov, and V.~Vanchurin, ``{Vector Inflation},''
  \emph{JCAP}, vol.~06, p. 009, 2008.

\bibitem{BeltranJimenez:2008iye}
J.~Beltran~Jimenez and A.~L. Maroto, ``{A cosmic vector for dark energy},''
  \emph{Phys. Rev. D}, vol.~78, p. 063005, 2008.

\bibitem{Libanov:2016kfc}
M.~Libanov, S.~Mironov, and V.~Rubakov, ``{Generalized Galileons: instabilities
  of bouncing and Genesis cosmologies and modified Genesis},'' \emph{JCAP},
  vol.~08, p. 037, 2016.

\bibitem{Kobayashi:2016xpl}
T.~Kobayashi, ``{Generic instabilities of nonsingular cosmologies in Horndeski
  theory: A no-go theorem},'' \emph{Phys. Rev. D}, vol.~94, no.~4, p. 043511,
  2016.

\bibitem{Evslin:2011vh}
J.~Evslin and T.~Qiu, ``{Closed Timelike Curves in the Galileon Model},''
  \emph{JHEP}, vol.~11, p. 032, 2011.

\bibitem{Easson:2011zy}
D.~A. Easson, I.~Sawicki, and A.~Vikman, ``{G-Bounce},'' \emph{JCAP}, vol.~11,
  p. 021, 2011.

\bibitem{Sawicki:2012pz}
I.~Sawicki and A.~Vikman, ``{Hidden Negative Energies in Strongly Accelerated
  Universes},'' \emph{Phys. Rev. D}, vol.~87, no.~6, p. 067301, 2013.

\bibitem{Rubakov:2016zah}
V.~A. Rubakov, ``{More about wormholes in generalized Galileon theories},''
  \emph{Theor. Math. Phys.}, vol. 188, no.~2, pp. 1253--1258, 2016.

\bibitem{Kolevatov:2016ppi}
R.~Kolevatov and S.~Mironov, ``{Cosmological bounces and Lorentzian wormholes
  in Galileon theories with an extra scalar field},'' \emph{Phys. Rev. D},
  vol.~94, no.~12, p. 123516, 2016.

\bibitem{Mironov:2019fop}
S.~Mironov, ``{Mathematical Formulation of the No-Go Theorem in Horndeski
  Theory},'' \emph{Universe}, vol.~5, no.~2, p.~52, 2019.

\bibitem{Akama:2017jsa}
S.~Akama and T.~Kobayashi, ``{Generalized multi-Galileons, covariantized new
  terms, and the no-go theorem for nonsingular cosmologies},'' \emph{Phys. Rev.
  D}, vol.~95, no.~6, p. 064011, 2017.

\bibitem{Cai:2017dyi}
Y.~Cai and Y.-S. Piao, ``{A covariant Lagrangian for stable nonsingular
  bounce},'' \emph{JHEP}, vol.~09, p. 027, 2017.

\bibitem{Creminelli:2016zwa}
P.~Creminelli, D.~Pirtskhalava, L.~Santoni, and E.~Trincherini, ``{Stability of
  Geodesically Complete Cosmologies},'' \emph{JCAP}, vol.~11, p. 047, 2016.

\bibitem{Cai:2016thi}
Y.~Cai, Y.~Wan, H.-G. Li, T.~Qiu, and Y.-S. Piao, ``{The Effective Field Theory
  of nonsingular cosmology},'' \emph{JHEP}, vol.~01, p. 090, 2017.

\bibitem{Cai:2017tku}
Y.~Cai, H.-G. Li, T.~Qiu, and Y.-S. Piao, ``{The Effective Field Theory of
  nonsingular cosmology: II},'' \emph{Eur. Phys. J. C}, vol.~77, no.~6, p. 369,
  2017.

\bibitem{Kolevatov:2017voe}
R.~Kolevatov, S.~Mironov, N.~Sukhov, and V.~Volkova, ``{Cosmological bounce and
  Genesis beyond Horndeski},'' \emph{JCAP}, vol.~08, p. 038, 2017.

\bibitem{Ageeva:2021yik}
Y.~Ageeva, P.~Petrov, and V.~Rubakov, ``{Nonsingular cosmological models with
  strong gravity in the past},'' \emph{Phys. Rev. D}, vol. 104, no.~6, p.
  063530, 2021.

\bibitem{Mironov:2022quk}
S.~Mironov and A.~Shtennikova, ``{Stable cosmological solutions in Horndeski
  theory},'' \emph{JCAP}, vol.~06, p. 037, 2023.

\bibitem{Mironov:2023wxn}
S.~Mironov and M.~Valencia-Villegas, ``{Stability of nonsingular cosmologies in
  Galileon models with torsion: A no-go theorem for eternal subluminality},''
  \emph{Phys. Rev. D}, vol. 109, no.~4, p. 044073, 2024.

\bibitem{Mironov:2024ffx}
------, ``{Healthy Horndeski cosmologies with torsion},'' \emph{JCAP}, vol.~07,
  p. 030, 2024.

\bibitem{Kaluza:1921tu}
T.~Kaluza, ``{Zum Unit\"atsproblem der Physik},'' \emph{Sitzungsber. Preuss.
  Akad. Wiss. Berlin (Math. Phys. )}, vol. 1921, pp. 966--972, 1921.

\bibitem{Vainshtein:1972sx}
A.~I. Vainshtein, ``{To the problem of nonvanishing gravitation mass},''
  \emph{Phys. Lett. B}, vol.~39, pp. 393--394, 1972.

\bibitem{Babichev:2013usa}
E.~Babichev and C.~Deffayet, ``{An introduction to the Vainshtein mechanism},''
  \emph{Class. Quant. Grav.}, vol.~30, p. 184001, 2013.

\bibitem{Koyama:2013paa}
K.~Koyama, G.~Niz, and G.~Tasinato, ``{Effective theory for the Vainshtein
  mechanism from the Horndeski action},'' \emph{Phys. Rev. D}, vol.~88, p.
  021502, 2013.

\bibitem{Kobayashi:2019hrl}
T.~Kobayashi, ``{Horndeski theory and beyond: a review},'' \emph{Rept. Prog.
  Phys.}, vol.~82, no.~8, p. 086901, 2019.

\bibitem{Fernandes:2022zrq}
P.~G.~S. Fernandes, P.~Carrilho, T.~Clifton, and D.~J. Mulryne, ``{The 4D
  Einstein\textendash{}Gauss\textendash{}Bonnet theory of gravity: a review},''
  \emph{Class. Quant. Grav.}, vol.~39, no.~6, p. 063001, 2022.

\bibitem{Padilla:2012dx}
A.~Padilla and V.~Sivanesan, ``{Covariant multi-galileons and their
  generalisation},'' \emph{JHEP}, vol.~04, p. 032, 2013.

\end{thebibliography}

\end{document}